\RequirePackage{fix-cm}

\documentclass[smallcondensed]{svjour3}
\journalname{Data Mining and Knowledge Discovery}

\smartqed  % flush right qed marks, e.g. at end of proof
\usepackage[numbers]{natbib}
\usepackage{graphicx}
\usepackage{color}
\graphicspath{{images/}{./}}
\usepackage[caption=false,font=footnotesize]{subfig}
\usepackage[cmex10]{amsmath}
\usepackage{amssymb}
\usepackage{amsopn}
\usepackage{bm}
\usepackage{multirow}
\usepackage{pifont}
\usepackage{algorithm}
\usepackage[noend]{algpseudocode}

\usepackage{url}

\newcommand{\cQ}{\mathcal{Q}}

\newcommand{\cB}{\mathcal{B}}
\newcommand{\cV}{\mathcal{V}}
\newcommand{\cE}{\mathcal{E}}
\newcommand{\cG}{\mathcal{G}}
\newcommand{\tilG}{\widetilde{\mathcal{G}}}
\newcommand{\tilE}{\widetilde{\mathcal{E}}}
\newcommand{\cW}{\mathcal{W}}
\newcommand{\cM}{\mathcal{M}}

\newcommand{\cD}{\mathcal{D}}

\newcommand{\cmark}{\ding{51}}
\newcommand{\xmark}{-}

\usepackage{xspace}
\newcommand{\ALGO}{D$^3$TS\xspace}
\newcommand{\probname}{selective harvesting\xspace}
\newcommand{\Probname}{Selective harvesting\xspace}
\newcommand{\ProbName}{Selective Harvesting\xspace}

\newcommand{\fm}[1]{#1}

\algrenewcomment[1]{\(\triangleright\) #1}
\algnewcommand{\LineComment}[1]{\State \(\triangleright\) #1}

\usepackage{array}
\newcolumntype{L}[1]{>{\raggedright\let\newline\\\arraybackslash\hspace{0pt}}m{#1}}

\begin{document}

\title{Selective Harvesting over Networks}



\author{Fabricio Murai \and
        Diogo Renn\'o \and \\
        Bruno Ribeiro \and
        Gisele L.\ Pappa \and \\
        Don Towsley \and
        Krista Gile}


\institute{\scriptsize
           F.\ Murai, D.\ Renn\'o \and G. L.\ Pappa \at
           Universidade Federal de Minas Gerais, Brazil \\
                \email{\{murai,renno,glpappa\}@dcc.ufmg.br}           %  \\
           \and
           B. Ribeiro \at
           %Department of Computer Science,
           Purdue University \\
              \email{ribeiro@cs.purdue.edu}           %  \\
              \and
D. Towsley \and K. Gile \at
           %College of Information and Computer Sciences,
           University of Massachusetts Amherst \\
              \email{towsley@cs.umass.edu; gile@math.umass.edu}           %  \\
}

\date{Received: date / Accepted: date}

\maketitle
%\vspace{-0.5cm}
\begin{abstract}
\fm{Active search on graphs} focuses on collecting certain labeled nodes (targets)
given global knowledge of the network topology and its edge weights
(encoding pairwise similarities) under a query budget constraint. However, in
most current networks, nodes, network topology, network size, and edge weights
are all initially unknown. In this work we introduce {\em \probname}, a
variant of active search where the next node to be queried must be chosen among
the neighbors of the current queried node set; the available training data for
deciding which node to query is restricted to the subgraph induced by the
queried set (and their node attributes) and their neighbors (without any node
or edge attributes).
%The queried set is expanded by querying neighbors of nodes
%in the set.
Therefore, \probname is a sequential decision problem, where we
must decide which node to query at each step. A classifier
trained in this scenario can suffer from what we call a {\em tunnel vision}
effect: without any recourse to independent sampling,
the urge to only query
promising nodes forces classifiers to gather increasingly biased training data,
which we show significantly hurts the performance of active search methods and
standard classifiers. We \fm{demonstrate} that it is possible to collect a much larger
set of targets by using multiple classifiers, not by combining their
predictions as a weighted ensemble, but switching between classifiers used at
each step, as a way to ease the tunnel vision effect. We \fm{discover} that switching
classifiers collects more targets by (a) diversifying the training data
and (b) broadening the choices of nodes that can be queried in the future.
This highlights an {\em exploration, exploitation, and diversification}
trade-off in our problem that goes beyond the exploration and exploitation
duality found in classic sequential decision problems. Based on these
observations we propose \ALGO, a method based on multi-armed bandits for
non-stationary stochastic processes that enforces classifier diversity, which
outperforms all competing methods on five real network datasets in our
evaluation and exhibits comparable performance on the other two.
\end{abstract}

%!TEX root = main.tex

\section{Introduction}

%GLP - thinking about it network search is not a good name.

\fm{Active search on graphs}~\cite{Garnett:2011wt,Ma:2015ut,Wang2013} is a technique
for finding the largest number of {\em target nodes} -- i.e., nodes with a
certain label -- in a network by querying nodes in a weighted graph, under a
query budget constraint. Nodes have hidden labels but the network topology and
edge weights are {\bf fully observable} and {\bf any node} can be queried at
any time. Edge weights encode some form of node similarity that can be used to
improve querying efficiency. Unfortunately, edge weights, network topology and
node information are rarely available to be downloaded from one centralized
place (except by the network's owner, if any). As a result, today's
prevalent method to collect network data is to query neighbors of already
queried nodes (crawling). Like \fm{active search on graphs}, other similar
techniques such as learning to crawl~\cite{Gouriten2014,Pant2005}, also assume
that edge weights between the queried nodes and their neighbors are observed.
But in a variety of network crawling problems, such as crawling online social
networks, (micro) blog networks, and citation networks, a node query often
reveals only node attributes. This process poses an entirely new set of
challenges for active search and other similar methods.

\fm{In this paper we introduce {\em \probname},
where the goal is the same as in active search, but instead of assuming
that the network topology is given,
our node querying is subject to a partial and evolving understanding of the network
More precisely,
the knowledge about the network is restricted to the set of
queried nodes and their connections to the rest of the network.} \Probname
starts from a seed node (typically a target) and proceeds by
querying nodes from the {\bf border set}, i.e.\ neighbors of already queried
nodes. \Probname generalizes active sampling, a similar task where node attributes are not observed \cite{pfeiffer2012}.
By leveraging information contained in these attributes, \probname algorithms can attain better performance in applications of active sampling, such as (i) identifying students involved in academic dishonesty at a college/university; (ii) investigating securities fraud and (iii) identifying students who smoke/drink for intervention purposes. In these cases, target nodes are
\fm{individuals that have a given trait.}
%Problems with partially observed networks include, for example, user recruitment
%problems.
%FM While existing heuristics for active search
%FM \cite{Garnett:2011wt,Wang2013,Ma:2015ut} can be adapted to partially 
%FM observed networks, they assume homophily. In scenarios where homophily is not too strong, alternative approaches need to be considered. We propose an approach based on
%FM learning a classifier, which has the potential to excel when provided with enough data,
%FM especially if homophily assumptions are violated. 
%FM Yet, training a classifier for {\em selective harvesting} is challenging due
%FM to detrimental ``tunnel vision'' effects: the classifier is fit to
%FM observations that depend on previous selection choices of the same classifier,
%FM the hidden network topology, and the distribution of node labels over the
%FM network. The locality of crawling forbids any true form of sample diversity or
%FM independence that can be used to break this tunnel effect. In such scenarios
%FM a na\"ive learning method performs quite poorly.%, unfortunately.
%FM %traditional active search methods perform quite poorly, unfortunately. 

Training a classifier for {\em \probname} is a challenging task due to
the fact that the classifier must be trained over 
observations that depend on previous choices of the same classifier,
the hidden network topology, and the distribution of node features over the
network. We call this the {\em tunnel vision effect}. 
Unlike active search, {\em \probname} has no recourse
to true randomness or sample independence that can ease the tunnel effect.
Under partially observed networks,
traditional active search methods perform quite poorly.%, unfortunately.

%We discover that it is possible to collect a much larger set of target nodes by
%using multiple classifiers, not by combining their predictions as a weighted
%ensemble, but switching between classifiers used at each step, as a way to ease
%the tunnel vision effect. We show that switching classifiers collects more
%target nodes by
We discover that it is possible to collect a much larger set of target nodes by
\fm{using a round robin scheme, which switches
between different types of classifiers (e.g., Logistic Regression, Random Forests) when}
predicting labels in different steps. We show that this strategy collects more
target nodes by 
(a) diversifying the training data and (b) broadening the
choices of nodes that can be queried in the future.
Based on these
observations, we propose Directed Diversity Dynamic Thompson Sampling (\ALGO),
a Multi-Armed Bandit (MAB) algorithm for non-stationary stochastic processes
that intelligently selects a classifier at each step to decide which neighbor
to query. This is in sharp contrast with ensemble techniques, which combine predictions
from several classifiers at each step. We show that these techniques (e.g., bagging
and boosting) do not perform as well as \ALGO due to the tunnel vision effect.

Unlike typical MAB problems, %including restless bandits~\cite{whittle1988restless},%
 where there is a clear exploration and exploitation tradeoff,
the standard MAB approach, which forces convergence to the ``best classifier'', would be suboptimal
in the presence of the tunnel vision effect. 
This gives rise to what we refer as {\em exploration, exploitation, and diversification} tradeoff.
\fm{\ALGO aims to induce continual diversification w.r.t.\ training data and potential node choices by using multiple
distinct classifiers, which plays a similar role to sample independence and eases the tunnel vision effect.}

\begin{figure}
\centering
\includegraphics[width=0.6\columnwidth,height=2.1in]{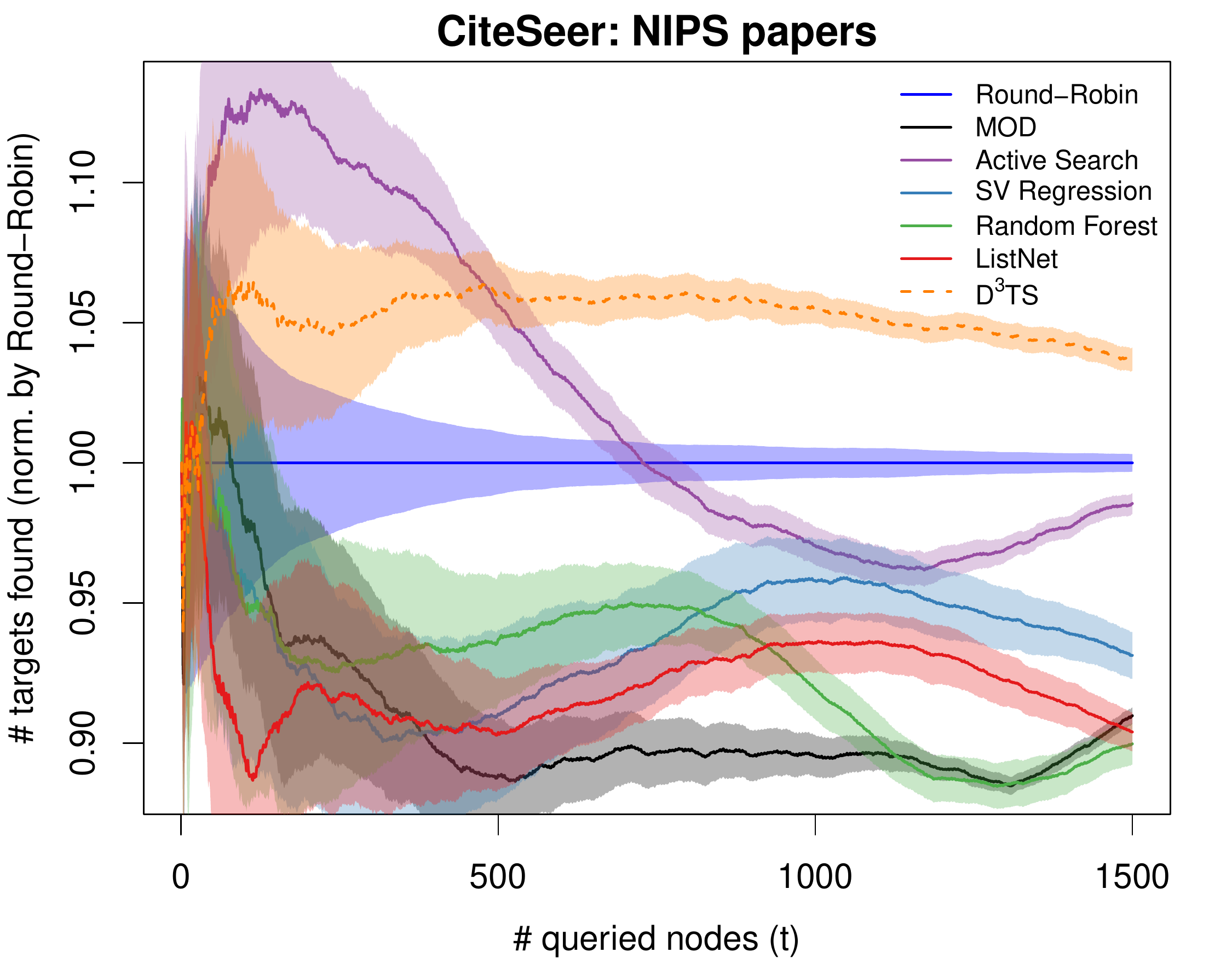}
\caption{Lines show the (scaled) average number of targets found by round-robin, five na\"ive classifiers and D$^3$TS against the total number of queries $(t)$.
Shadows indicate 95\% confidence intervals over 80 runs, each starting at a seed uniformly chosen from target population.
Surprisingly, round-robin use of five classifiers (including poor-performing ones) outperforms any single classifier in the CiteSeer network. 
We also see that the best-performing active search method (Wang et al.~\cite{Wang2013}) has its relative accuracy eroded over time (and we will see why this is likely due to the {\em tunnel vision effect}). We include the proposed method (\ALGO) results, which are consistently better than all competing methods for $t \geq 500$.
 }
\label{fig:intro}
\end{figure}
Interestingly, we find that even a round-robin selection of five distinct
classifiers often performs better than just using the best classifier or the
best active search method for each dataset.
Consider simulation results shown in Figure~\ref{fig:intro}
(the simulation is further explained in Section~\ref{sec:datasets}, for now we focus only on the overall results). 
Figure~\ref{fig:intro} shows the number of queries (x-axis) against the number of target nodes found in the CiteSeer paper co-citation network (NIPS papers as targets) normalized by the number of target nodes found by a round robin selection of five distinct simple classifiers (y-axis); the details of these simple classifiers are given in Section~\ref{sec:search}.
Note that over time the cumulative gain of the best active search method for this dataset (Wang et al.~\cite{Wang2013}) slowly erodes until it is worse than the na\"ive round-robin approach. 
Our analysis shows that this erosion can be attributed to the tunnel vision effect.
Each of the five simple classifiers when used on their own are consistently outperformed by the round-robin approach, and the best such classifiers also suffer from a performance erosion over time.
In contrast, our proposed method, %Directed Diversity Dynamic Thompson Sampling (\ALGO),
\ALGO, consistently and significantly outperforms state-of-the-art methods, the round-robin approach, and na\"ive approaches. % (even when the latter are applied in hindsight).
%
%\subsection{Contributions}
The contributions of this work are as follows:
\begin{enumerate}
\item {\bf Formulation and characterization of \ProbName and Classifier Diversity:} We introduce \probname and show that state-of-the-art methods such as
active sampling~\cite{pfeiffer2012,Bnaya:2013ck} and active search~\cite{Garnett:2011wt,Wang2013,Ma:2015ut} perform poorly in these settings. 
\fm{We show that switching between various classifiers is helpful to achieve greater performance. This works not because we are exploring classifiers in order to find the best one or because we are combining their predictions as an ensemble. Instead, the use of multiple classifiers -- helps improve accuracy in two complementary ways.}
It achieves {\em border set diversity}, by exploring regions and thus avoiding remaining in a region where target nodes have been depleted. It also achieves {\em training sample diversity}, where diverse classifiers create enough diversity of observations to ease the {\em tunnel vision effect}.
%We also show that uniformly sampling nodes for diversity has serious detrimental effects in performance.

\item {\bf Directed Diversity Dynamic Thompson Sampling (\ALGO)}: we propose
  \ALGO, a method for \probname which combines different classifiers,
  and show that it consistently outperforms state-of-the-art methods. We
  evaluate the proposed framework on several real-world networks and observe
  that \ALGO outperforms all tested methods on five out of seven datasets and
  exhibits similar performance on the other two.\footnote{\fm{The software and scripts to reproduce results presented in this work are available as an R package \url{http://bitbucket.com/after-acceptance}. All the data used in this work is publicly available from different sources.}}
\end{enumerate}

{\bf Outline.} In \S\ref{sec:formulation} we formalize the \probname problem
and present a generic algorithm for solving it. In \S\ref{sec:search} we
describe existing and potential approaches to solve this problem and show that
the tunnel vision effect hurts their performance. In \S\ref{sec:diversity} we
investigate why classifier diversity -- i.e., using multiple classifiers -- can
mitigate the tunnel vision effect. We propose D$^3$TS in \S\ref{sec:d3ts}.
Datasets and results of our evaluation are described in
\S\ref{sec:methodology}. Related work is described in~\S\ref{sec:related}.
\fm{In \S\ref{sec:discussion} we discuss alternatives to the proposed method and explain
why they cannot be applied or why they do not perform well.} Last, our conclusions are presented
in \S\ref{sec:conclusions}.

%!TEX root = main.tex

\section{Problem Formulation} \label{sec:formulation}

\begin{figure}[!t]
\centering
\includegraphics[width=2.0in]{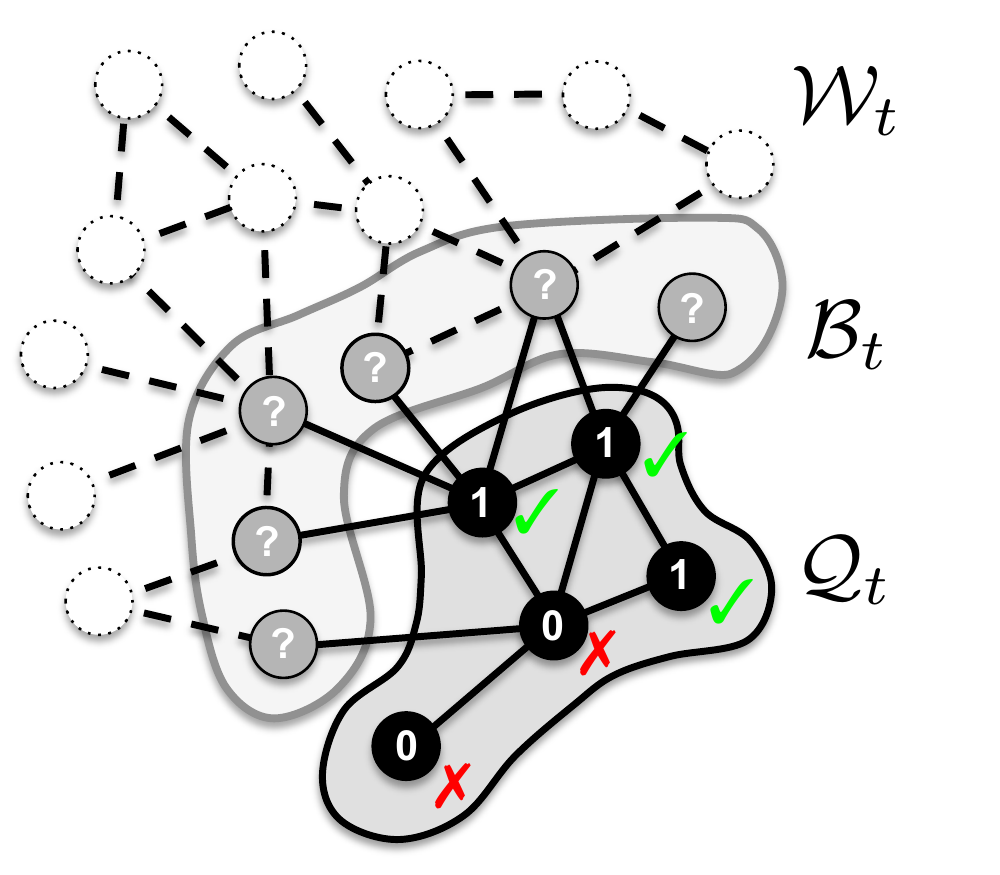}
\vspace{-10pt}
\caption{Representation of the search state over an unknown graph $\cG$ after $t = 4$ steps.
Solid nodes and edges show the subgraph $\tilG_t$.
Black nodes represent queried nodes.
Unknown labels of nodes in $\cB_t$ are represented by a question mark ``?''. }
\label{fig:notation}
\end{figure}

In this section we formalize the \probname problem and introduce notation used throughout this work.
Let $\cG=(\cV,\cE)$ denote an undirected graph representing the network topology.
%where $n=|V|$ is the number of nodes.
Each node $v \in \cV$ has $M$ attributes (domain-related properties of the
nodes) encoded without loss of generality as an attribute vector $\mathbf{a}_v
\in \mathbb{R}^M$.

\fm{In active search problems,} the goal is to find a large set of nodes in $\cV$
that satisfy a given search criterion
(e.g., nodes that exhibit a given attribute) under the constraint that no more than $T$ nodes
can be queried.
The search criterion is a boolean function $f: \cV \rightarrow \{0,1\}$.
Formally, let $\cV_+ \subset \cV$ be the set of all target nodes,
i.e.\ all $v$ such that $f(v)=1$. We define node labels
$y_v$ as
\[
  y_v = f(v) = \left\{ \begin{array}{ll}
          1 & \quad\text{ if $v \in \cV_+$,}\\
          0 & \quad\text{ otherwise.} 
        \end{array} \right. \quad \forall v \in \cV
\]

\Probname is a variant of active search. In active search, the topology is assumed to be known.
In \probname, the search is subject to a limited but evolving knowledge of the network.
This knowledge is expanded by querying nodes in $\cV$, which reveals their labels, neighbors and attribute vectors.
A set of pre-queried nodes $\cQ_0 \subset \cV$ is given as input (typically consisting of one target node).
Subsequent queries are restricted to neighbors of already queried nodes.

At any step $t$, nodes belong to one of three sets:
$\cQ_t$, the set of previously queried nodes; 
$\cB_t$, the set of neighbors of queried nodes that have not been queried (referred as border nodes or border set);
or $\cW_t$, the set of unobserved nodes, which are invisible to the algorithm. Figure~\ref{fig:notation} illustrates a snapshot of the search process (see caption for details).

Let $\tilG_t = (\cQ_t, \tilE_t)$ denote the subgraph of $\cG$ given by the
subgraph induced by nodes in $\cQ_t \cup \cB_t$ minus edges in the subgraph
induced by $\cB_t$ (i.e., $\tilG_t$ contains all edges between nodes in $\cQ_t$
plus edges connecting $\cQ_t$ to $\cB_t$).
The graph $\tilG_t$ is the portion of the network visible at step $t$.
%A queried node $v \in \cQ(t)$ has observed label $y_v$, which is either $1$ or $0$.
In $\tilG_t$, label $y_v$ is only known for nodes in $\cQ_t$.

%The notation is illustrated in Fig.~\ref{fig:notation}. 

{\bf Generic solution.} Given an {\em initial input graph} $\tilG_0$, an
algorithm for \probname must decide at each step $t=1,\ldots,T$ what {\em action}
to take, i.e., what border node $v \in \cB_t$ to query, given the currently
available network information. This {\em action} returns $v$'s label,
attributes and connections, which is included as {\em additional input} to the
search in step $t+1$. Node $v$ label (0 or 1) can be thought of as the {\em
payoff} obtained by querying that node.  The algorithm's {\em output} is the
list of target nodes found in $T$ steps. The best algorithm is the one that
yields the largest total payoff, i.e., yields the largest number of target
nodes.

\section{Background}\label{sec:search}

\begin{table}[t]
\centering
\label{tab:litreview}
\begin{tabular}{lcccccc}
  & PNB \cite{pfeiffer2012} &  SN-UCB1 \cite{Bnaya:2013ck} & MOD \cite{Avrachenkov:2014tk} &  AS \cite{Wang2013} & \ALGO (ours)\tabularnewline
\hline
Unknown network  & \cmark & \cmark &  \cmark & \xmark & \cmark 
\vspace{2pt}
\tabularnewline
Uses node features  & \xmark & \xmark &  \xmark & \xmark & \cmark 
\vspace{2pt}
\tabularnewline
Unknown\ neighbor  & \multirow{2}{*}{\xmark} & \multirow{2}{*}{\xmark} & \multirow{2}{*}{\xmark}  & \multirow{2}{*}{\xmark} & \multirow{2}{*}{\cmark} \tabularnewline
attributes  &  &  &  &  &
\vspace{2pt}
\tabularnewline
Fits model to evol- & \multirow{2}{*}{\xmark} &  \multirow{2}{*}{\cmark} & \multirow{2}{*}{\xmark}   & \multirow{2}{*}{\xmark} &  \multirow{2}{*}{\cmark} \tabularnewline
\vspace{2pt}
ving observations  &  &  &  &  & \tabularnewline
Scalable & \xmark & \cmark & \cmark  & \cmark & \cmark \tabularnewline
\hline 
\end{tabular}
\caption{Comparison of heuristics for \probname: Active Sampling (PNB), Social Network UCB1 (SN-UCB1), Maximum Observed Degree (MOD), 
and Active Search (AS).}
\end{table}

In this section, we review methods for searching networks that can be used for or adapted to \probname.
These methods exploit correlation between labels of connected nodes to find targets.
In addition, we review statistical models that could be used as an alternative (data-driven) approach.
In contrast to existing methods, this approach can leverage node attributes by training a statistical model to infer the
node's label from the observed graph.
As a slight abuse of terminology, we may refer to existing methods and base learners generically
as {\bf classifiers}, since both are used to infer border nodes' labels.

%In active search problems -- where the network topology is fully observable --
%previously proposed works \cite{Wang2013,Ma:2015ut} define index policies
%that try to account for the future impact of querying a node. More precisely, 
%such policies choose the node $v$ that maximizes, at each step $t+1$,
%the sum of the expected payoff $\mu_t(v)$ and the estimate $s_t(v)$ of the query's future impact
%discounted by a factor $\alpha_{t+1}$: 
%\begin{equation}
%v_{t+1} = \arg \max_{v \in \cB_t} \mu_t(v) + \alpha_{t+1} s_t(v).
%\end{equation}
%%
%The future impact $s_t(v)$ is typically defined as the sum of terms, one for each non-queried node. Most of these
%nodes are unknown in \probname (i.e., belong to $W_t$). Hence, it is not clear how to specify the future impact $s_t(v)$ in \probname;
%such specification is beyond the scope of this paper.

%Existing approaches to active search are based on homophily or, more generally, some
%smooth function over the graph. In what follows, we describe adaptations of current methods to \probname.

%In what follows, we describe how existing methods can be used or adapted to \probname.
%All these methods are heuristics that assume homophily with respect to node labels to find target nodes.
%In addition, we consider a data-driven approach capable of leveraging node attributes, where a statistical model is trained to infer the
%node's label from the observed graph.

\subsection{Existing methods}

A few works in the literature provide methods that can be used for or adapted to \probname.
A subclass of \probname methods known as active sampling~\cite{pfeiffer2012,Bnaya:2013ck}
does not account for node attributes. Our problem is closely related to the
graph-theoretic myopic budgeted online covering
problem~\cite{Avrachenkov:2014tk,Khuller:2013iv,Borgs:2012jl}. In this problem,
all nodes are relevant (equivalently, all nodes are targets) and the task is to
find a connected set of nodes that yields the largest cover (i.e., the largest
set $\cQ_t \cup \cB_t$ set). The closest problem to ours is that addressed by
\fm{active search on graphs}~\cite{Garnett:2011wt,Wang2013,Ma:2015ut}, where
nodes have hidden labels but the topology and edge weights are fully observed and any
node can be queried at any time. Algorithms for myopic budgeted online covering and active search can
be adapted for \probname; active sampling methods require little or no modification.

We adapt four representative methods of the above to \probname: Active Sampling~\cite{pfeiffer2012} (PNB -- in reference to the authors surnames), Maximum Observed Degree (MOD)~\cite{Avrachenkov:2014tk}, Social Network UCB1 (SN-UCB1)~\cite{Bnaya:2013ck},
and Active Search (AS)~\cite{Wang2013}. Table~\ref{tab:litreview}
summarizes the key differences between these methods and the proposed method, \ALGO.

%Among heuristics that are readily applicable to \probname we consider the Active Sampling method in \cite{pfeiffer2012} and the SN-UCB1 \cite{Bnaya:2013ck} algorithm.
%Among heuristics that can be adapted, we consider the Active Search method in \cite{Wang2013}. None of these methods can make use of node features.

{\bf Active Sampling (PNB)}: PNB is a representative algorithm from the class of active sampling approaches proposed in \cite{pfeiffer2012}.
PNB estimates a border node's payoff value $y_v$ using a weighted average of the payoffs
of observed nodes two hops away from $v$, where weights are the number of common neighbors with $v$.
Border nodes are included among these observed nodes, requiring all payoffs to be collectively estimated by a label propagation
procedure based on %computationally expensive
Gibbs Sampling.
%This excessive computational cost shows in the running (wall-clock) time of the algorithm (see XXX).
PNB also tracks a running average of payoff values
acquired from random jumps, which we do not allow in our simulations since these are not possible in \probname.
Please see \cite{pfeiffer2012} for a detailed description of PNB's parameters.%
%\footnote{We do not include the Active Exploration method~\cite{Pfeiffer:2013cikm} in
%our study because it assumes that border nodes' attributes
%are directly observable.}.

{\bf Social Network UCB1 (SN-UCB1)}:  The SN-UCB1 search algorithm proposed in~\cite{Bnaya:2013ck} divides border nodes into
equivalence classes and samples from theses classes using a multi-armed bandit algorithm.
Equivalence classes are composed of all border nodes
 connected to the same set of queried nodes. These classes are volatile: they split, disappear and appear over time,
requiring the use of a variant of the UCB1 called VUCB1. Although this method
learns about the equivalence classes, it does not learn a statistical model
that can account for node attributes.
Similar to \probname, it assumes partial but evolving knowledge about the network.

{\bf Maximum Observed Degree (MOD)}: MOD is a myopic algorithm proposed in \cite{Avrachenkov:2014tk} to maximize the network cover as it explores a graph.
MOD is the optimal greedy cover algorithm in a finite random power law network (under the Configuration Model \cite{newman2003structure}) with degree distribution coefficient either one or two.
In our simulations we adapt MOD to select the border node with the maximum number of target neighbors in the queried set (ties are resolved randomly).
From the expected excess degree results in \cite{Avrachenkov:2014tk} such border nodes are rich with target neighbors provided that the underlying network exhibits strong
homophily with respect to node labels.

{\bf Active Search}:  this method, proposed by Wang et al.~\cite{Wang2013}, attempts to find target nodes
by assuming that labels are defined by a smooth function over the graph edges. To
estimate the unknown labels, it attaches to each labeled instance a virtual node containing the instance's label
and then performs label propagation on the original graph. It assumes that the graph
is known, which allows it to estimate the future impact of choosing a given border node.
We adapt Active Search to run label propagation only on the observed graph.% and disregard
%the calculation of impact on unobserved nodes%
%We select Wang et al.~\cite{Wang2013}, one of the most recent and practically-oriented active search approaches and restrict the set of nodes that
%can be queried to the border set for comparison.
%Our results show that Wang et al.\ can perform poorly in some graphs while \ALGO consistently outperforms Wang et al.\  in all but one dataset (Wikipedia).
\footnote{Although the method proposed by Wang et al.~\cite{Wang2013} is outperformed by a more recent proposal~\cite{Ma:2015ut} in active search problems, we found the opposite to be true when the graph is not fully observable. In addition to being highly sensitive to the parameterization, the most recent method computes and stores a dense correlation matrix between all visible nodes, which is hard to scale beyond $10^5$ nodes.}

\subsection{Data-driven methods}

A data-driven \probname algorithm trains a statistical model to estimate the expected payoff $\mu_t(v)$
obtained from querying border node $v \in \cB_t$, based on $v$'s relationship with the observed graph $\tilG_t$ at step $t$. We encode this relationship
as a ``local'' feature vector $\mathbf{x}_{v|\tilG_t}$, which we describe next. Note that $v$'s features differ from $v$'s attributes (denoted by $\mathbf{a}_v$).
Since $v$'s attributes are not observable until it is queried, we compute $v$'s local features from the observed graph $\tilG_t$
to use as training data for base learners.

\subsubsection*{Feature Design}\label{sec:features}

We define features for each border node in $v \in \cB_t$. They are divided into:

 \begin{itemize}
  \item {\bf Pure structural features}: observed degree and number of triangles formed with observed neighbors.
\item {\bf Structure-and-attribute blends}: number and fraction of target neighbors, number and fraction of triangles
  formed with two non-target (and with two target) neighbors, number and fraction of neighbors mostly surrounded
  by target nodes, fraction of neighbors that exhibit each node attribute, probability of finding a target exactly after two random walk steps from
  border node.\footnote{Other seemingly obvious features (e.g., number of non-target neighbors) are not considered due to colinearity. Longer random walk paths are too expensive to be used in most real networks.}
\end{itemize}

We build upon features typically used in the literature \cite{robins2007introduction,robins2007recent}.
We also use a Random Walk (RW) transient distribution to build features: we consider the expected payoff observed by a RW that departs from node $u \in \cB_t$ and performs two steps, given by
\begin{equation}
  x_{u|\tilG_t}^{\textrm{(RW)}} = \frac{\sum_{(u,v) \in \tilE_t}\sum_{ (v,w) \in \tilE_t, w \in \cQ_t} y_{w}}{C_{u|\tilG_t}} \label{eq:rw}
\end{equation}
where $C_{u|\tilG_t}$ is the number of such paths of length two in $\tilG_t$.
Note that the RW is not restricted to the immediate neighbors of $u$. 
Also, this is not an average among the nodes two hops away from $u$; this feature depends on the connectedness of the border node's neighborhood
in the observed graph.

\subsubsection*{Base Learners}

The feature vector described above can be given as input to any learning method able to generate a ranking of border nodes. We consider classification, regression and ranking methods as suitable candidates for this task.
The classification representatives include {\bf Logistic Regression} and {\bf Random Forests},
because they provide ways to rank border nodes according to how confident the
model is that each border node is a target. {\bf Exponentially Weighted Least
Squares} (EWLS) and {\bf Support Vector Regression} are included by modeling
the task as a regression problem, and the list-wise learning-to-rank method
{\bf ListNet} \cite{Cao:2007en} for directly outputting ranks.
%We also include {\bf Bagging} and {\bf AdaBoost} as representatives of classifier ensembles.
%A sophisticated learner specifically tailored for networks, Network Regression
%\cite{stojanova2012}, assumes that the network is fully observable and that
%edge weights can be computed from node attributes and, therefore, is not
%considered in this work.
We briefly describe EWLS and ListNet below and refer the reader to
\cite{friedman2009} for descriptions of other methods.

{\bf Exponentially Weighted Least Squares (EWLS)}: computes
weights $\mathbf{w}$ that, given a forgetting factor $0 \ll \beta \leq 1$ and regularization parameter $\lambda$,
minimize the loss function
\[
\sum_{i=1}^t \beta^{t-i} |y_t - \mathbf{x_t}^\top \mathbf{w}|^2 + \beta^t \lambda \| w \|^2.
\]
EWLS gives more weight to recent observations. The weights $\mathbf{w}$ are suitable for
fast online updates \cite[Section~4.2]{liu2011kernel}. Setting $\beta=1$ reduces EWLS to $\ell_2$-regularized Linear Regression.

{\bf ListNet}: This is a representative method from the list-wise approaches for learning to rank
(a Machine Learning task where the goal is
to learn how to rank objects according to their relevance to a query) \cite{Cao:2007en}. It assumes that
the observed ranking $\boldsymbol \pi$ is a random variable that depends on the objects' scores (where $\pi_1$ is the top-ranked object).
The scores are determined by a neural network that is trained by minimizing the
K-L divergence between the probability distribution over $ \boldsymbol{\hat \pi}$ and the probability distribution
over a ranking $\boldsymbol \pi$ derived from ground-truth scores. In our context, $P(\boldsymbol \pi)$ is given by
\[
P(\boldsymbol \pi = \langle \pi_1,...,\pi_{|\cB_t|} \rangle) =
\prod_{i=1}^{|\cB_t|} \left[ \exp(y_{\pi_i})/\sum_{j=i}^{|\cB_t|} \exp(y_{\pi_j}) \right].
\]
Since the goal is not to predict the object-wise relevance, all of the statistical
power of this method goes into learning the ranking.

As with any learning approach, in the ``small data'' regime (few observations collected)
a base learner may perform worse than heuristic methods that assume
homophily w.r.t.\ node labels. To mitigate issues related to fitting a learner to few observations and
yet allow a fair comparison with the heuristic methods, we
query the first $20$ nodes using MOD.\footnote{In comparison to other combinations of length and
heuristic used in the ``cold start'' phase, this was found to work best.}

%Methods and datasets are described in Sections~\ref{sec:methods} and \ref{sec:datasets}.

%Henceforth, we refer to methods in either approach generically as {\em classifiers}.
%We observe that the success of the two approaches varies across datasets.
%In fact, there is no classifier that always performs best even among those belonging to the same approach.
%But more importantly, a classifier that exhibits the best performance on one network
%might perform poorly on another. Unfortunately, without prior information about the network,
%there is no principled way to choose the right classifier.
%In what follows, we discuss a na\"ive way to mitigate this problem and make
%an interesting finding.

% Activate the following line by filling in the right side. If for example the name of the root file is Main.tex, write
% "...root = Main.tex" if the chapter file is in the same directory, and "...root = ../Main.tex" if the chapter is in a subdirectory.
 
%!TEX root =  main.tex

\section{Tunnel Vision and the power of Classifier Diversity}\label{sec:diversity}

In \probname the goal is to find the most number of target nodes with a limited
query budget. This requires methods to try to sample only promising target
nodes, which causes a given classifier to gather increasingly biased training
data, a phenomenon that we call {\em tunnel vision effect}. Unfortunately, it
is unlikely that we can find a method which provably compensates for this bias
in our training data, $\cQ_t$. Even if we query
border nodes randomly at each step, we cannot determine the probability of seeing
any given node in the border set $\cB_t$, as this
would require assessing the probability of all possible sample paths from the
given seed nodes, which includes paths containing nodes not yet observed, i.e.,
nodes in $\cW_t$ in Figure~\ref{fig:notation}, an unfeasible task as we {\bf do
not} know the network topology. This is likely why active search and base
learners by their own do not work well for \probname tasks. This is also why
importance weighted sampling~\cite{beygelzimer2009} cannot be used to remove
the bias in these tasks.

\begin{table*}[t]
\centering
\begin{tabular}{lrrrrrrr}
  \hline
   \multirow{3}{*}{{\bf Methods}} & \multicolumn{7}{c}{{\bf Datasets} (budget $T$)}\\
   \cline{2-8}
                     & \multicolumn{1}{c}{{\bf CS}} & \multicolumn{1}{c}{{\bf DBP}}& \multicolumn{1}{c}{{\bf WK}} & \multicolumn{1}{c}{{\bf DC}} & \multicolumn{1}{c}{{\bf KS}} & \multicolumn{1}{c}{{\bf DBL}} & \multicolumn{1}{c}{{\bf LJ}} \\ 
%     & (1400)    & (1000) &  (500) & (125)    & (1500) & (1250) & (1250) \\
    & (1500)    & (700) &  (400) & (100)    & (700) & (1200) & (1200) \\
\hline
 PNB & {\bf 833.2$^*$} & 260.6$^*$ & 107.7$^*$ &  24.3$^*$ & 178.3$^*$ & 599.5$^*$ & 632.4$^*$ \\ 
  SN-UCB1 & 568.9$^*$ & 272.3$^*$ &  71.8$^*$ &  23.2$^*$ & 133.2$^*$ & 399.1$^*$ & 573.7$^*$ \\ 
  MOD \cmark & 746.8$^*$ & 403.0$^*$ & 140.9$^*$ &  35.7$^*$ & 159.6$^*$ & 580.3$^*$ & 584.1$^*$ \\ 
  Active Search \cmark & 808.9$^*$ & 412.2$^*$ & {\bf 143.4}$\ \,$ &  22.6$^*$ & 215.3$^*$ & 684.9$^*$ & 654.2$^*$ \\ 
   \hline
Logistic Regression & 764.5$^*$ & 452.5$\ \,$ &  86.2$^*$ &  35.8$\ \,$ & 122.1$^*$ & {\bf 744.4}$\ \,$ & 732.0$\ \,$ \\
  Random Forest \cmark & 738.5$^*$ & 454.0$^*$ & 127.2$^*$ &  37.2$\ \,$ & 215.6$^*$ & 725.4$\ \,$ & 728.3$^*$ \\ 
  EWLS & 808.2$^*$ & {\bf 462.4}$\ \,$ &  82.5$^*$ &  35.2$^*$ & 142.3$^*$ & 656.9$^*$ & 694.4$^*$ \\ 
  SV Regression  \cmark & 770.6$^*$ & 456.3$^*$ &  85.0$^*$ & {\bf  37.6}$\ \,$ & 205.3$^*$ & {\bf 757.1$^*$} & 736.1$\ \,$ \\ 
  ListNet \cmark & 742.0$^*$ & 448.0$^*$ &  92.5$^*$ &  34.4$^*$ & 146.3$^*$ & 730.7$\ \,$ & {\bf 742.8}$\ \,$ \\ 
  %Bagging & 745.6$^*$ & 445.0$^*$ & 99.1$^*$ & 34.7$^*$ & 224.9$^*$ & -- & -- \\ 
   %AdaBoost & 751.5$^*$ & 443.5$^*$ & 98.0$^*$ & 34.5$^*$ & 218.4$^*$ & 669.9 & -- \\  
   \hline
Round-Robin (all \cmark) & 822.2$^*$ & 454.5$^*$ & 135.3$^*$ &  37.3$\ \,$ & {\bf 234.9$^*$} & 696.0$^*$ & 716.0$^*$ \\ 
  \ALGO (all \cmark) & {\bf 851.2}$\ \,$ & {\bf 464.0}$\ \,$ & {\bf 144.7}$\ \,$ & {\bf  37.9}$\ \,$ & {\bf 247.6}$\ \,$ & 729.5$\ \,$ & {\bf 737.3}$\ \,$ \\
  \hline
% PNB & {\bf 833.2$^*$} & 260.6$^*$ & 107.7$^*$ &  24.3$^*$ & 178.3$^*$ & 599.5$^*$ & 632.4$^*$ \\ 
%  SN-UCB1 & 568.9$^*$ & 272.3$^*$ &  71.8$^*$ &  23.2$^*$ & 133.2$^*$ & 399.1$^*$ & 573.7$^*$ \\ 
%  MOD \cmark & 746.8$^*$ & 403.0$^*$ & 140.9$^*$ &  35.7$^*$ & 159.6$^*$ & 580.3$^*$ & 584.1$^*$ \\ 
%  Active Search \cmark & 808.9$^*$ & 412.2$^*$ & {\bf 143.4} &  22.6$^*$ & 215.3$^*$ & 684.9$^*$ & 654.2$^*$ \\ 
%   \hline
%Logistic Regression \cmark & 764.5$^*$ & 452.5 &  86.2$^*$ &  35.8 & 122.1$^*$ & {\bf 744.4} & 732.0 \\ 
%  EWLS & 808.2$^*$ & {\bf 462.4} &  82.5$^*$ &  35.2$^*$ & 142.3$^*$ & 656.9$^*$ & 694.4$^*$ \\ 
%  SV Regression & 770.6$^*$ & 456.3$^*$ &  85.0$^*$ &  37.6 & 205.3$^*$ & {\bf 757.1$^+$} & 736.1 \\ 
%  Random Forest \cmark & 757.7$^*$ & 439.9$^*$ & 116.7$^*$ &  34.6$^*$ & 214.2$^*$ & 718.1 & 729.0$^*$ \\ 
%  ListNet \cmark & 742.0$^*$ & 448.0$^*$ &  92.5$^*$ &  34.4$^*$ & 146.3$^*$ & 730.7 & {\bf 742.8} \\ 
%   \hline
%Round-Robin & 829.9$^*$ & 455.0$^*$ & 139.8$^*$ & {\bf  37.9} & {\bf 239.3} & 699.2$^*$ & 712.7$^*$ \\ 
%  \ALGO  & {\bf 854.9} & {\bf 464.9} & {\bf 144.6} & {\bf  37.7} & {\bf 240.5} & 730.5 & {\bf 738.0} \\ 
%  \hline
Target population size     & 1583$\quad$           & 725$\quad$            & 202$\quad$            & 56$\quad$            & 1457$\quad$           & 7556$\quad$           & 1441$\quad$           \\ \hline
\end{tabular}
\caption{Average number of targets found by each method after $T$ queries based on 80 runs.
{\bf Datasets.} {\bf CS}: CiteSeer, {\bf DBP}: DBpedia, {\bf WK}: Wikipedia, {\bf DC}: DonorsChoose,
{\bf DBL}: DBLP, {\bf KS}: Kickstarter and {\bf LJ}: LiveJournal. Budget $T$ is respectively set to number of targets $\times 1, \times 1, \times 2, \times 2, \times \frac{1}{2}, \times \frac{1}{6}, \times \frac{5}{6}$ truncated to hundreds. First four rows correspond to existing methods; five subsequent rows are base learners.
Round-Robin and \ALGO combine methods indicated by (\cmark).
Means whose difference to \ALGO's is statistically significant at the 95\% confidence level are indicated by ($^*$). 
Best two results on each dataset are shown in bold.
{\bf Parameters.} PNB: same as in \cite{pfeiffer2012}; Active Search: same as in \cite{Wang2013};
ELWS: $\beta=.99$, $\lambda=1.0$; Logistic Regression and SV Regression: penalty $C$ set using fast heuristic implemented in 
R package \texttt{LiblineaR} \cite{LiblineaR:2015}; Random Forest: no.\ variables $ = \sqrt{\textrm{no. features}}$, number of trees $=100$ (DBL and LJ use classical decision trees for speed, others use conditional inference trees \cite{hothorn2006unbiased}); ListNet: no.\ iterations $=100$, tolerance $=10^{-5}$.} 
\label{tab:best}
\end{table*}

%\begin{table}[ht]
%\centering
%\begin{tabular}{rllll}
%  \hline
%                 &  Best\\ 
%  \hline
%CiteSeer & PNB () \\
%DBpedia & \\
%Wikipedia & \\
%Donors & \\
%Kickstarter & \\
%DBLP & \\
%LiveJournal
%
%   \hline
%\end{tabular}
%\caption{CiteSeer: NIPS} 
%\end{table}

To demonstrate the tunnel vision effect and show how classifier diversity can mitigate it,
we conduct a large set of simulations.
 We simulate searches using four heuristics -- MOD, PNB, Social Network-UCB1 (SN-UCB1) and Active Search, five base learners -- Logistic Regression, Exponentially Weighted Least Squares (EWLS), Support Vector Regression, Random Forest and ListNet %, and two ensemble methods -- Bagging and AdaBoost --
on seven networks and summarize the results in Table~\ref{tab:best} (network datasets and target populations are described in Section~\ref{sec:datasets}).
\fm{We observe that the best classifier varies across datasets. More surprisingly, the best classifier for one dataset may be the worst
for another (see Active Search on Wikipedia and on DonorsChoose).}

We then consider a set of classifiers $\mathcal{M}$ that typically exhibit good performance
and cycle between them during the search, in a Round-Robin (RR) fashion.
Based
on Table~\ref{tab:best}, we pick $\mathcal{M}$=\{MOD, Active Search, Support Vector Regression, Random Forest, ListNet\}.\footnote{We choose MOD in lieu of PNB because MOD is orders of magnitude faster.
Among the base learners, we choose one representative of regression (SV Regression), classification (Random Forest) and ranking (ListNet) methods.} We use this set of classifiers {\bf throughout the rest of this paper}, unless otherwise noted.
One might
expect RR's performance to be the average
of the performance results yielded by the standalone counterparts, but this is not the case. Interestingly, switching classifiers at each step
outperforms the best classifier in $\cM$ on the CiteSeer and Kickstarter datasets,
and finds at least 92\% as many target nodes as the best classifier on other datasets.
In what follows we investigate why the use of multiple classifiers can improve \probname's performance.

\subsection{Leveraging diversity through the use of multiple classifiers}

%Predicting RR's performance
%is hard because \probname is dynamic in many aspects: the border set --
%which contains the set of possible payoffs at each step -- changes as the search proceeds;
%querying different nodes reveals different parts of the graph and
%impacts a classifier in different ways. Since classifiers can disagree about
%which node to query next, RR will produce a different sample path from those generated by individual classifiers. 

\begin{figure}
\centering
\includegraphics[height=0.23\textheight]{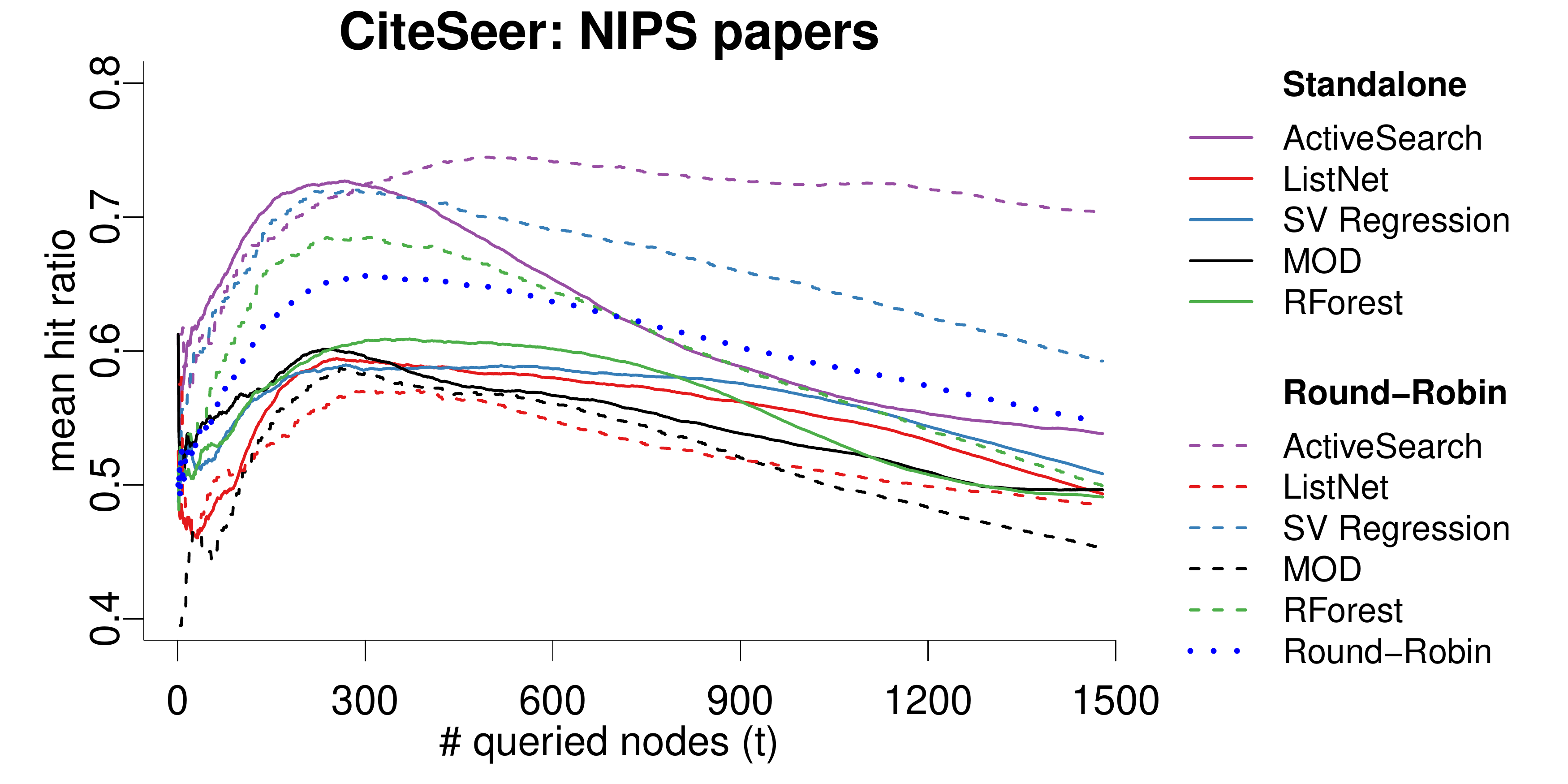}
\caption{Round-robin can have higher hit ratios for each of its classifiers than their standalone counterparts.}
\label{fig:hitratio}
\end{figure}

We observe that RR outperforms all five classifiers in $\cM$ on CiteSeer (Table~\ref{tab:best}). Consequently, at least one of them must perform better under RR
than on its own. In order to identify which ones do, we show in Figure~\ref{fig:hitratio} the hit ratio -- number
of target nodes found divided by number of queries performed using each classifier up to time $t$ -- under RR
and when used by itself,
averaged over 80 runs. Interestingly, after $t=400$ all classifiers exhibit similar (relative difference $\leq 10\%$) or better performance under RR than when used alone. 

We
propose two hypotheses to explain this performance improvement:
\begin{enumerate}
\item[(a)] {\bf Border Hypothesis}: RR explores regions of the graph containing more targets that are likely to be scored high by a classifier,  i.e.\ RR infuses diversity in the border set. 
% \item[(a)] {\bf Hypothesis Border}: RR explores other regions of the graph where we have not yet 
%depleted the target nodes, i.e.\ RR infuses diversity in the border set. 
\item[(b)] {\bf Training Hypothesis}: 
Observations from different classifiers can be used to train the others to generalize better and cope with self-reinforcing sampling biases,
i.e., diversity in the training set produces a classifier that is better at finding target nodes.
\end{enumerate}
Note that these hypotheses are not mutually exclusive.
%In fact, there is likely to be interaction between border set diversity and training diversity. 
In what follows, we perform controlled simulations to isolate and study each hypothesis.

Training set diversity directly impacts model parameters.
Model parameters, in turn, determine how the border set will change.
Therefore, to assess the impact of training set diversity we must hold the border set diversity constant and vice-versa.
This is the key idea behind the two controlled sets of simulations described next.
To perform them, we instrumented our simulator to load, from another simulation run,
(i) the feature vector $\mathbf{x}_{\sigma_t|\tilG_t}$ of node $\sigma_t$ queried in step $t$, and label $y_{\sigma_t}$, and
(ii) the observed graph $\tilG_t$ at each step $t$. 
In what follows, we show the results obtained using the support vector regression (SVR) model. We denote node $\sigma_t$'s feature vector and label 
simply by $\mathbf{x}_{t}$ and $y_t$, respectively, to make it easier to follow.

\begin{figure*}
  \center
    \subfloat[Simulations for studying Border Hypothesis]{\includegraphics[width=0.78\textwidth]{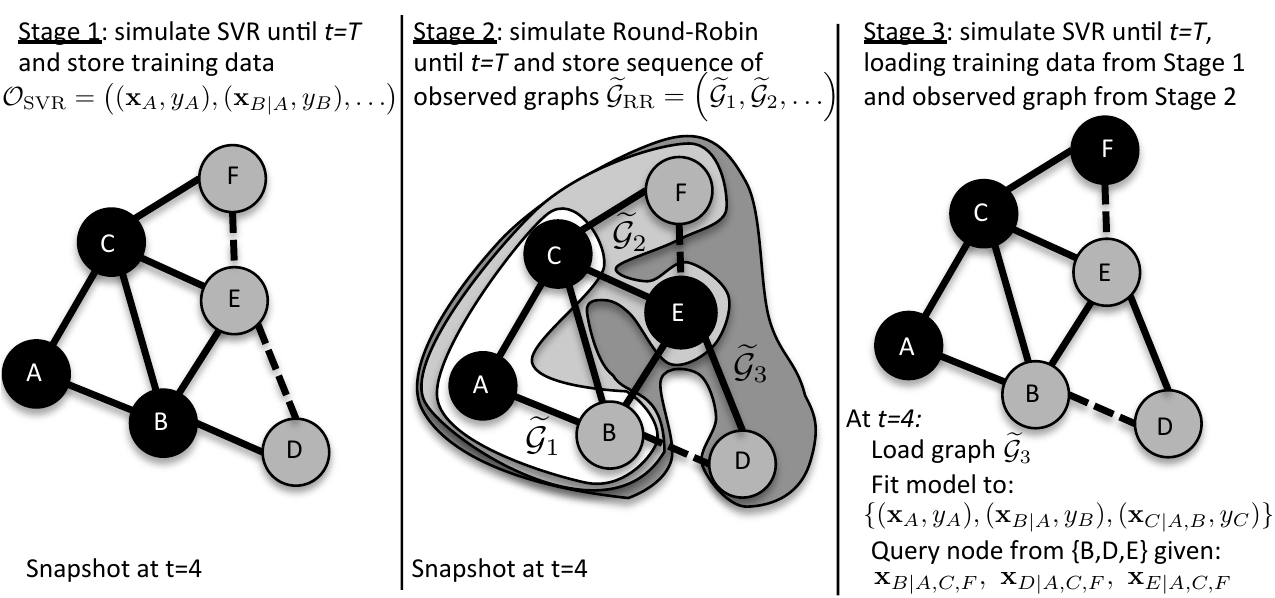}
     \label{fig:exp1}}\\
        \subfloat[Simulations for studying Training Hypothesis]{\includegraphics[width=0.78\textwidth]{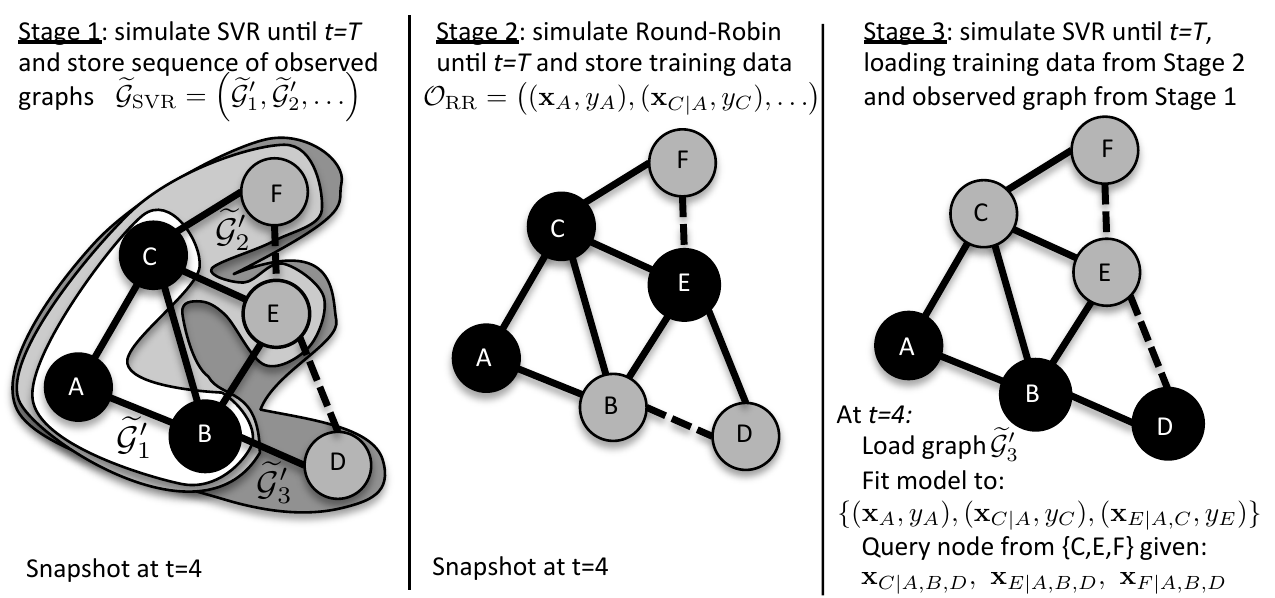}
     \label{fig:exp2}}
\label{fig:experiments}
\caption{(a) We study the Border Hypothesis by recreating the sequence of SVR models from the original
simulation run (stage 1) and using them to query nodes on a sequence of observed graphs collected using round-robin (stage 2). (b) We study the Training Hypothesis by recreating the sequence of observed graphs from the original
simulation run (stage 1) and using a SVR trained on the samples collected using round-robin (stage 2) to query nodes.}
\end{figure*}     

%\begin{figure*}
%\centering
%\includegraphics[width=0.8\textwidth]{images/experiment1}
%\caption{Experiment to test Border Hypothesis by recreating the sequence of SVR models from the original
%simulation run (stage 1) and using them to query nodes on a sequence of observed graphs collected using round-robin (stage 2).}
%\label{fig:exp1}
%\end{figure*}

%\subsection{Studying hypothesis Border}

{\bf Border Hypothesis.} Our experiment consists of three stages (Fig.~\ref{fig:exp1}). First, we store the sequence of observations
(i.e., pairs feature vector, label)
$\mathcal{O}_\mathrm{SVR}=\left((\mathbf{x}_{1},y_{1}),\ \ldots, (\mathbf{x}_{T},y_{T})\right)$ corresponding to nodes queried when
searching a network dataset $\mathcal{D}$ using SVR. Second, we store the sequence
of observed graphs $\tilG_\mathrm{RR} = \left( \tilG_1, \ldots, \tilG_T \right)$ when searching $\mathcal{D}$ by cycling between models in the
set $\mathcal{M}$.
Last, we simulate another SVR-based search on $\mathcal{D}$, loading the observed graph at each time step $t$
from $\tilG_\mathrm{RR}$. However, instead of training the SVR model with
observations collected on that run (which most likely differ from those collected during the first stage),
we gradually feed it with observations from $\mathcal{O}_\mathrm{SVR}$, one for each simulation step $t$. Therefore, we will reproduce the
sequence of classifiers from the first stage, but subject to a different sequence of observed graphs.

%\subsection{Studying hypothesis Training}

{\bf Training Hypothesis.} As before, our experiment consists of three stages (Fig.~\ref{fig:exp2}). In the first stage, we store the sequence of observed graphs $\tilG_\mathrm{SVR} = \left( \tilG^\prime_1, \ldots, \tilG^\prime_T \right)$ when searching $\mathcal{D}$
using a SVR model. Second, we store the sequence of observations
$\mathcal{O}_\mathrm{RR}=\left((\mathbf{x}_{1}^\prime,y_{1}^\prime), \ldots, (\mathbf{x}_{T}^\prime,y_{T}^\prime)\right)$ collected when searching $\mathcal{D}$
by cycling among classifiers in $\mathcal{M}$. Last, we simulate another SVR-based search, loading the observed graph at each time step $t$ from $\tilG_\mathrm{SVR}$,
but feeding it observations from $\mathcal{O}_\mathrm{RR}$, one by one. Hence, the classifier is fit to a different set of observations,
but the search is subject to the same sample path as the SVR-based search from the first stage.

\begin{figure}
\centering
\includegraphics[width=0.45\columnwidth]{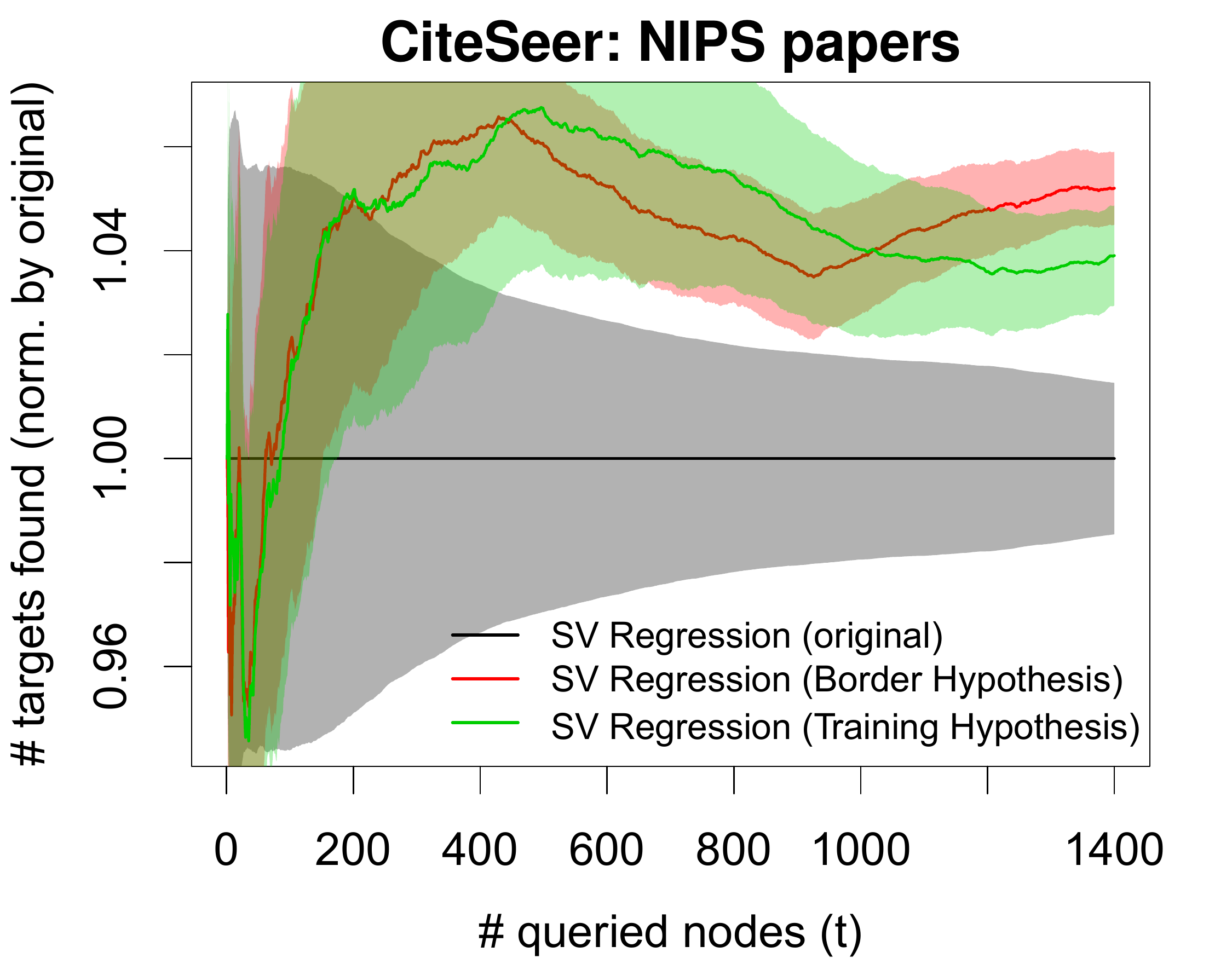}%{images/citeseer_hypotheses}
\caption{SVR classifier and two ways to ease the {\em tunnel vision effect}: {\bf border set diversity} and {\bf training set diversity} improve performance by ensuring greater diversity in query choices and by diversifying the training data, respectively.}
\label{fig:hypotheses}
\end{figure}

Figure~\ref{fig:hypotheses} contrasts the average number of target nodes found by the original SVR-based search on CiteSeer
against those obtained in each set of simulations based on 80 runs.
The 95\% confidence intervals for the mean at $t=700$ are $[393.8,413.1]$,
$[416.6,427.5]$ and $[417.1,436.7]$. These statistics corroborate the hypotheses that the border set and the training data collected by
the round-robin policy contribute to improving the performance of the SVR model.
%Although the plot suggests
%that the training set diversity has more impact on the performance than the border set diversity, in non-controlled simulations we cannot distinguish the
%two factors since they interact. 

Intuitively, when a base learner is fit to the nodes it queried, it tends to
specialize in one region of the feature space and the search consequently only
explores similar parts of the graph, which can severely undermine its potential
to find target nodes. One way to mitigate this overspecialization would be to
sample nodes from the border set probabilistically, as opposed to
deterministically querying the node with the highest score. This alternative is
investigated in Appendix~\ref{app:random}, where the ranking associated with
each classifier is mapped into a probability distribution. The results show no
significant performance improvement over those obtained when a single
classifier chooses nodes to query deterministically.

The round-robin policy infuses diversity in the training set
without sacrificing performance.
%This differs from the notion of diversity implied in the active learning literature \cite{} as it
%does not imply random sampling.
This diversity is achieved by
``asking another classifier'' what is the best node to query at a given step. In scenarios where all classifiers
would have performed reasonably well if used alone, learning from another's classifier query is likely to improve
one classifier's ability to find targets, especially when they disagree.

Yet, different classifiers inherently exhibit different performances on a dataset. \fm{Clearly, we want to choose more accurate classifiers more often, but in order to do so, three challenges must be addressed:
\begin{enumerate}
\item We do not know a priori which classifiers are more accurate on a dataset;
\item Classifiers' accuracy varies as their parameters are updated and the border set changes;
\item Continual exploration must be ensured, since converging to an arm would make the search more susceptible to the tunnel vision effect.
\end{enumerate}
Challenge (1) is typically addressed by Multi-Armed Bandit (MAB) algorithms. Challenge (2) constrains the set of possible MAB algorithms to those designed for MAB problems with non-stationary reward distributions. Challenge (3) is specific to selective harvesting (the exploration-exploitation-diversification trade-off).
In the following section, we propose a method that addresses all these challenges. We call it Directed Diversity Dynamic Thompson Sampling because it is based
on the Dynamic Thompson Sampling algorithm for MAB problems and because it leverages
diversity in a ``directed way'' as opposed to randomly sampling nodes.}

%In the following section,
%we propose a method that learns these inherent performances online and thus improves upon round-robin
%by not using all classifiers an equal number of times.
%We call it Directed Diversity Dynamic Thompson Sampling because it is based
%on the Dynamic Thompson Sampling algorithm for multi-armed bandit problems and because it leverages
%diversity in a ``directed way'' as opposed to randomly sampling nodes.

%!TEX root = main.tex

\section{Directed Diversity Dynamic Thompson Sampling (D$^3$TS)}\label{sec:d3ts}

%explain how the problem could be modeled as MAB
%Multi-armed Bandits (MABs) are sequential decision problems where a forecaster has to choose, at each step,
%an action from a set of available actions. After choosing an action, an observable payoff is
%obtained. The goal is to design a policy that maximizes the total payoff. The literature on MABs
%is extremely vast and contains many specializations of the more general formulation. As a result,
%there are several ways of mapping the problem of searching a network using models as experts that
%provide advice on which nodes to choose. In this section, we describe the most appropriate mappings,
%which we use in the results section.
\fm{This section is divided in two parts.
First, we discuss the relationship between \probname and multi-armed bandits.
Then, in the light of this discussion, we propose the \ALGO algorithm.}

\subsection{Relationship between \Probname and Multi-Armed Bandits}

\Probname with multiple classifiers can be cast as a Multi-Armed Bandit (MAB) problem. In a MAB problem, a forecaster is given the number of arms $K$ and the number of rounds $T$. For each round $t$, nature generates a payoff vector $\mathbf{r}_t = (r_{1,t},\ldots,r_{K,t}) \in [0,1]^K$ unobservable to the forecaster.\footnote{In general, rewards can be normalized to be in $[0,1]$.} The forecaster chooses an arm $I_t \in \{1,\ldots,K\}$ and receives payoff $r_{I_t,t}$, with the other payoffs hidden. The goal is to maximize the cumulative payoff obtained.
MAB problems can be classified according to how the payoff vector is generated. In {\bf stochastic bandit problems}, each entry $r_{i,t}$ in the payoff vector is sampled 
independently, from an unknown distribution $\nu_i$, regardless of $t$. In {\bf adversarial bandit problems}, the payoff vector $\mathbf{r}_t$ is chosen by an adversary which, at time $t$, knows the past, but not $I_t$. Stochastic and adversarial
bandits do not cover the entire problem space, as the payoff vector distribution may vary over time in a less arbitrary way than in adversarial bandits. In {\em stochastic bandit problems with non-stationary distributions} or {\em dynamic bandit problems}, the mean payoff vector can evolve according to random shocks or change at pre-determined points in time. MAB problems may also include context, which provides the forecaster with side information about the optimal action at a given step. In {\em contextual bandits}, a context $\mathbf{x}_{a,t}$ is drawn (from some unknown probability distribution) for each action $a \in \mathcal{A}_t$ available in step $t$. The context may be provided explicitly or through recommendations of a set of experts.

In \probname, the sequential decision problem consists of choosing the node to query at each step, given recommendations from several models.
There are two ways of mapping \probname to a MAB problem. The first (and simplest) mapping is context-free. Each model is represented by an arm (i.e., the problem reduces to one of choosing a model at each time step). Models are treated as black boxes that will ``internally'' query a node and return the node's label.
The queried node's label is seen as the model's payoff. The second mapping falls into the class of contextual bandits. Each border node represents an action and each model represents an expert that provides recommendations on how to choose the actions. Node features correspond to action contexts, which are used by the experts to compute their recommendations.

Despite the potential advantage of accounting for node features directly and combining the advice of several models,
most algorithms for contextual bandits assume fixed and small (relative to the time horizon) sets of actions, whereas
the border set is dynamic and potentially orders of magnitude larger than the query budget. Among context-free bandits,
we claim that algorithms for stochastic bandits with non-stationary distributions are the best candidates for combining
classifiers in \probname, as we observe that the average hit ratio can drift over time (Fig.~\ref{fig:hitratio}).
While adversarial bandits allow payoff distributions to change arbitrarily, they cannot exploit the fact that
the mean payoff evolves in a well-behaved manner. A thorough comparison of several bandit algorithms described in
Appendix~\ref{app:mabs} supports our claim. Our comparison includes the Exp4 and Exp4.P algorithms for contextual bandits,
which combine the prediction of all classifiers in a similar way that traditional ensemble methods do.

 \begin{algorithm}[t]
 % \caption{Pay-to-Recruit(initial set $V_0$, initial heuristic $\mathcal{M} = \{\mathcal{M}_0\}$, budget $T$)}
  \caption{\ALGO(budget $T$, model set $\mathcal{M}$, threshold $C \geq 2$)}
  \label{alg:d3ts}
  \begin{algorithmic}[1]
    \LineComment{Assume $\mathcal{B}_t$ is updated after each iteration.}
%  \For{$r \in V_0$}
%  \ForAll{neighbor $v$ of $r$ in $G_{\mathcal{R}}(t)$}
%      \State $\mathbf{x}_v \gets $computeFeatures$(v, y_r, \chi_r, N_r)$
%    \EndFor
%  \EndFor
  \For{$t$ in $1,\ldots,T$}
  \For{$k$ in $1,\ldots,|\mathcal{M}|$}
  \State $\hat{r}_t^{(k)} \sim \mbox{Beta}(\alpha_k,\beta_k)$
  \EndFor
  \State $I_t = {\arg\max}_{k \in 1,\ldots,K} \hat{r}_t^{(k)} $\label{model}
  \State $\hat{\mathbf{y}} = $ estimate payoffs using classifier $I_t$ and $\tilG_t$  \label{payoff}
  \State $b = \arg\max_{v \in \mathcal{B}_t} \hat{y}_v $\label{max}
  \State $r_t = y_b = $ query$(b)$\label{acquire}
  \If{ $\alpha_{I_t} + \beta_{I_t} < C$ }
      \State $\alpha_{I_t} = \alpha_{I_t} + r_t$
      \State $\beta_{I_t} = \beta_{I_t} + (1-r_t)$
  \Else
        \State $\alpha_{I_t} = (\alpha_{I_t} + r_t)\times C/(C+1)$
      \State $\beta_{I_t} = (\beta_{I_t} + (1-r_t))\times C/(C+1)$
  \EndIf
  \State $\cM = $ update or retrain classifiers given new point $(x_{b|\tilG_t},y_b)$
  \EndFor
%\EndFunction
\end{algorithmic}
\end{algorithm}

\subsection{Proposed algorithm}

For the reasons above, we adapt the Dynamic Thompson Sampling (DTS) algorithm \cite{Gupta:2011df}
proposed for MABs with non-stationary distributions to the \probname problem.
DTS is based on the Thompson Sampling (TS) algorithm for stochastic MABs, where \fm{binary outcomes associated with each arm $k=1,\ldots,K$
are modeled as Bernoulli trials. The uncertainty on the probability parameter associated with arm $k$ is typically
modeled as a distribution $\mbox{Beta}(\alpha_k,\beta_k)$. The Beta distribution is the conjugate prior
for the Bernoulli distribution (thus providing computational savings on Bayesian updates).
TS performs exploration by choosing arms probabilistically, according to samples drawn from the corresponding distributions. More precisely,
at step $t$, TS samples $\hat{r}_t^{(k)} \sim \mbox{Beta}(\alpha_k,\beta_k)$
and selects the arm with the largest sample, i.e., $I_t = \arg\max_{k \in 1,\ldots,K} \hat{r}_t^{(k)}$.
Given the binary payoff $r_t$ received
after selecting arm $I_t$, the distribution parameters are updated according to the Bayesian rule, i.e.,
$\alpha_{I_t} = \alpha_{I_t} + r_t$ and $\beta_{I_t} = \beta_k + (1-r_t)$. In essence, DTS normalizes arm $k$'s parameters
such that $\alpha_k + \beta_k \leq C$, where $C$ is a bounding parameter.  We adapt DTS in two senses: (i) we combine DTS with the steps needed to perform search in \probname problems and (ii) we set the threshold $C$ to a much smaller value than the ones used in \cite{Gupta:2011df}, which allows us to incur more diversity.
This highlights an {\em exploration, exploitation and diversification} tradeoff
 in \probname that goes beyond the duality found in classic MAB problems, as simply converging to one arm would be suboptimal. The pseudo-code for
 D$^3$TS is shown in Algorithm~\ref{alg:d3ts}. In what follows we compare D$^3$TS against all approaches
 for \probname discussed in Section~\ref{sec:search}.}

%We experiment with representative algorithms of each of the following context-free subclasses:
%\begin{itemize}
%\item Stochastic Bandits: UCB1, Thompson Sampling (TS), $\epsilon$-greedy
%\item Adversarial Bandits: Exp3, Exp3.P \cite{Auer:2002hg} 
%\item Non-stationary stochastic bandits: Dynamic Thompson Sampling (DTS) \cite{Gupta:2011df}
%\end{itemize}

%The second mapping falls into the class of contextual bandits. Each node represents an action and each model represents an expert that gives recommendations on how to choose the actions. Node features correspond to action contexts, which are used by the experts to compute their recommendations. We experiment with Exp4 \cite{Auer:2002hg}  and Exp4.P \cite{Beygelzimer:2011wp}, which are representatives algorithms of this class.

% Activate the following line by filling in the right side. If for example the name of the root file is Main.tex, write
% "...root = Main.tex" if the chapter file is in the same directory, and "...root = ../Main.tex" if the chapter is in a subdirectory.
 
%!TEX root =  main.tex

%\section{Features, Methods and Datasets}

% Activate the following line by filling in the right side. If for example the name of the root file is Main.tex, write
% "...root = Main.tex" if the chapter file is in the same directory, and "...root = ../Main.tex" if the chapter is in a subdirectory.
 
%!TEX root = main.tex 
\section{Simulations}\label{sec:methodology}

This section describes the datasets used in our simulations, together with simulation results and comparisons with baseline methods.

\subsection{Datasets}\label{sec:datasets}

To evaluate the above search methods, we use seven datasets corresponding to undirected and unweighted networks containing node attributes.
In the following we describe each of the datasets summarized in Table~\ref{tab:description}. Basic statistics for each network are shown in Table~\ref{tab:stats}.

\begin{table*}[t]
  \centering
  %\scriptsize
  \begin{tabular}{lllll}
  \hline
  Dataset & nodes &  edges &  node attributes &  target nodes \tabularnewline
\hline
  %\hline 
DBpedia & places & hyperlinks & place type & admin. regions\tabularnewline
%\hline 
CiteSeer & papers & citations & venues & top venue \tabularnewline
%\hline 
Wikipedia & wikipages & links & topics & OOP pages \tabularnewline
%\hline 
Kickstarter & donors & co-donors & backed projects & DFA donors \tabularnewline
%\hline
DonorsChoose & donors & co-donors & awarded projects & $P$ donors \tabularnewline
%\hline 
LiveJournal & users & friendship & enrolled groups & top group\tabularnewline
%\hline 
DBLP & authors & co-authorship & conference & top conference \tabularnewline
\hline 
\end{tabular}
\caption{High-level description of each network.}
\label{tab:description}
%\vspace{-.05in}
\end{table*}
\begin{table*}
  \centering
  %\scriptsize
  \begin{tabular}{lcccc}
  \hline
  Dataset & $|\cV|$ & $|\cE|$ & $M$ & $|\cV_+|/|\cV|$\tabularnewline
\hline
DBpedia & 5.00K & 26.6K & 5  & 14.5\% \tabularnewline
CiteSeer &  14.1K & 42.0K & 10 & 13.1\% \tabularnewline
Wikipedia  & 5.27K & 64.6K & 93 & 3.83\% \tabularnewline
Kickstarter  &  27.8K & 2.77M  & 180  & 5.27\%  \tabularnewline
DonorsChoose & 1.15K & 6.60K & 284 & 4.96\% \tabularnewline
LiveJournal & 4.00M & 34.7M & 5K & 0.04\% \tabularnewline 
DBLP & 317K & 1.05M & 5K & 2.38\% \tabularnewline
%DBpedia & 5.00K & 26.6K & 5  & 14.5\% & 0.368 \tabularnewline
%CiteSeer &  14.1K & 42.0K & 10 & 13.1\% & 0.344\tabularnewline
%Wikipedia  & 5.27K & 64.6K & 93 & 3.83\% & 0.584\tabularnewline
%Kickstarter  &  27.8K & 2.77M  & 180  & 5.27\%  & 0.164\tabularnewline
%DonorsChoose & 1.15K & 6.60K & 284 & 4.96\% & 0.291\tabularnewline
%LiveJournal & 4.00M & 34.7M & 5K & 0.04\% & 0.620\tabularnewline 
%DBLP & 317K & 1.05M & 5K & 2.38\% & 0.648\tabularnewline
\hline 
\end{tabular}
\caption{Basic statistics of each network: $|\cV|$ (number of nodes), $|\cE|$ (number of edges), $M$ (number of attributes) and
$|\cV_+|/|\cV|$ (fraction of target nodes). %and $H$ (assortativity coefficient).
}
\label{tab:stats}
%\vspace{-.36in}
\end{table*}

The first three datasets have been used as benchmarks for Active Search \cite{Wang2013,Ma:2015ut}.
Despite the fact that Active Search assumes that the network topology is known, we can use these datasets to evaluate active search methods
by only revealing parts of the graph as the search proceeds. \fm{We define the target population
as in the Active Search work.}

{\bf DBpedia}: A network of 5000 populated places from the DBpedia
ontology formed by linking pairs whose corresponding Wikipedia pages
link to each other, in either direction. Places are marked
as ``administrative regions'', ``countries'', ``cities'', ``towns'' or ``villages''.
\fm{Target nodes are the ``administrative regions''.}

{\bf CiteSeer}: A paper citation network composed of the
top 10 venues in Computer Science. Papers are
annotated with publication venue. \fm{Target nodes
are the NIPS papers.}

{\bf Wikipedia}: A web-graph of wikipages
related to programming languages.
Pages are annotated with topics obtained
by thresholding a pre-computed topic vector \cite{Wang2013}.
\fm{Target nodes are webpages related to ``object oriented programming''.}

Two network datasets from the Stanford SNAP repository~\cite{leskovec2014snap} typically used to validate community detection algorithms
are also used. We label nodes belonging to the largest ground-truth community as targets. Other community memberships are used to
define a binary attribute vector $\mathbf{a}_v \in \{0,1\}^M$ for all $v \in V$.
%For the above three ``community datasets'',
%the target population is simply defined as
%the set of nodes belonging to the largest ground-truth community. Payoffs
%are set to $c^+ = 1$ if the node is a target and $c^- = -1$ otherwise. All networks are treated as undirected. 

 {\bf LiveJournal}: A blog community with OSN features, e.g.: users declare friendships and create
    groups that others can join. Users are annotated with
    the groups they joined.

 {\bf DBLP}: A scientific collaboration network
    where two authors are connected if they have
    published together. Authors are annotated with
    their respective publication venues.

Last, we use datasets containing donations to projects posted on two online crowdfunding websites.
To assess the performance of each classifier in low correlation settings, we build a social network connecting potential donors where edges are weak predictors of whether or not neighbors of a donor will also donate.
%From these weak edge signals \ALGO extracts a much stronger signal by considering the network local structure and node traits.
We label nodes as targets if they donated to a specific campaign. Historical donation data prior to that is used to build the network and define node attributes.

{\bf Kickstarter(.com)}: An online crowdfunding website. 
This dataset was collected by GitHub user {\em neight-allen} and consists of 3.04M donors that together made 5.87M donations to 87.3K projects.
%Because of the lack of a real social network on our crowdfunding dataset
We create a donor-to-donor network by connecting donors that donated to the same projects in the past.
More precisely, we assume that backers of small unsuccessful campaigns (between 100 and 600 backers) are all connected in a co-donation network -- say, their names are published on the campaign's website.
We choose campaigns with few donors so that the resulting network is sparse and the network discovery problem
challenges \ALGO.
   Our dataset has 180 small unsuccessful projects between 04/21/2009 and 05/06/2013,
   containing a total of 27.8K donors.
   We then choose the 2012 project (denoted DFA) that has the largest number of donors in our dataset.
   The goal of the recruiting algorithm is to recruit the 2012 DFA donors through the donor-to-donor network of past donations (2009--2011).
%We make one critical assumption, namely that there are no potential donors for the DFA project
 %  in the constructed network that did not donate just because they were unaware of the project.

{\bf DonorsChoose(.org)}: An online crowdfunding website where
    teachers of US public schools post classroom projects
    requesting donations (e.g., for a science project).
    The dataset is part of the KDD 2014 Cup containing 1.29M donors that together made 3.10M donations to 664K projects
    from 57K schools.
    Donations include information such as donor location, donation amount,
    awarded project, among other node features.
    As donors tend to be loyal to the same schools, we focus on the school that received the most donations in the dataset.
    We use projects from 2007 to 2012 to construct a donor-to-donor network 
    where an edge exists between two donors if they donated to the same project less than 48 hours
    apart. We then select the project $P$ in 2013 with the largest number
    of donations. 

%Table~\ref{tab:stats} shows statistics of the networks obtained from our datasets.
%The ratio $|\mathcal{T}|/|V|$ denotes the fraction of positive payoff nodes in the network and $H$
%is the assortativity coefficient~\cite{newman2002} to measure the amount of homophily in these networks w.r.t.\ node payoff values.
%It is worth noting that the task of finding positive payoff nodes is not easy.
%The positive payoff population varies from 0.04\% (LiveJournal) to 5.3\% (Kickstarter) of the total number of nodes in the analyzed networks.
%%The assortativity coefficient range is $H \in [-1,1]$, where $H=1$ when no edges exist between positive and negative payoff nodes,
%%$H=0$ when there is no homophily (edges disregard payoff values), and $H=-1$ if nodes with positive payoffs only have edges to negative payoff nodes.
%%Assortativity is essential to understand why one of our baseline methods works extremely well in some networks and not so well in others.

\subsection{Results}\label{sec:results}

\begin{table}[t]
\centering
\begin{tabular}{l|rr|rr|rr}
  \hline
  \multirow{2}{*}{Dataset}  & \multicolumn{2}{c|}{avg top 5} &\multicolumn{2}{c|}{avg top 3} & \multicolumn{2}{c}{avg top 1} \\
   \cline{2-7}
        & RR & D$^3$TS & RR & D$^3$TS & RR & D$^3$TS  \\
\hline
CiteSeer & 1.04 & 1.07 & 1.02 & 1.05 & 1.00 & 1.03 \\ 
  DBpedia & 1.01 & 1.03 & 1.00 & 1.02 & 0.98 & 1.01 \\ 
  Wikipedia & 1.16 & 1.20 & 1.05 & 1.08 & 0.97 & 1.01 \\ 
  DonorsChoose & 1.06 & 1.05 & 1.04 & 1.04 & 1.01 & 1.00 \\ 
  Kickstarter & 1.23 & 1.24 & 1.13 & 1.14 & 1.11 & 1.12 \\ 
  DBLP & 0.96 & 1.00 & 0.94 & 0.98 & 0.92 & 0.96 \\ 
  LiveJournal & 0.98 & 1.02 & 0.97 & 1.00 & 0.96 & 0.99 \\ 
  %\hline
%CiteSeer & 1.03 & 1.07 & 1.01 & 1.04 & 0.99 & 1.02 \\ 
%  DBpedia & 1.00 & 1.02 & 0.99 & 1.01 & 0.98 & 1.00 \\ 
%  Wikipedia & 1.11 & 1.19 & 0.99 & 1.05 & 0.94 & 1.01 \\ 
%  DonorsChoose & 1.03 & 1.04 & 1.01 & 1.03 & 0.99 & 1.01 \\ 
%  Kickstarter & 1.20 & 1.27 & 1.11 & 1.17 & 1.09 & 1.15 \\ 
%  DBLP & 0.96 & 1.00 & 0.94 & 0.98 & 0.92 & 0.96 \\ 
%  LiveJournal & 0.99 & 1.01 & 0.97 & 1.00 & 0.96 & 0.99 \\ 
\hline
\end{tabular}
\caption{Performance ratios: between RR (D$^3$TS) and average of top $k=1,3,5$ standalone classifiers.} 
\label{tab:mab}
\end{table}

In this section, we compare the performances of \ALGO, Round-Robin (RR) and standalone classifiers,
w.r.t.\ the number of targets found at several points in time.
We set the threshold $C=5$ in \ALGO and parameters of all classifiers as in Table~\ref{tab:best}.

We simulate \probname on each dataset for a large budget $T$, chosen in proportion to the target population size
(e.g., for DonorsChoose we set $T=100$, for Kickstarter we set $T=1500$).
In order to contrast RR's and \ALGO' performance against that obtained if side
information about the identity of the top $k$ performing classifiers on a given
dataset were available, Table~\ref{tab:mab} lists ratios between RR's (and
\ALGO') performance and the average performance of the top $k=1,3,5$ standalone
classifiers.
Note that we consider the top $k$ from all nine standalone classifiers described in Section~\ref{sec:search},
not only the classifiers used by RR (and \ALGO). Top classifiers vary across datasets.
%\footnote{We conducted some preliminary studies to identify dataset characteristics
%which favor the performance of some classifiers over others. A positive correlation between homphily (measured by the assortative coefficient~\cite{newman2002}) and Active Search's performance was observed, but a more detailed investigation is left as future work.}

Overall, we observe that RR's performance is comparable to that of the top 3 classifiers and can
sometimes outperform them (by up to 13\%). In the worst case, RR's performance is 92\% of that of the
best standalone classifier (DBLP). \ALGO consistently improves upon RR and yields
results at least as good as the best standalone classifier on all datasets except DBLP and LiveJournal, where its performance
is respectively 96\% and 99\% of that of the best classifier. \ALGO outperforms the best classifier by up to 15\% (Kickstarter).

\begin{figure*}
  \center
    \subfloat{\includegraphics[width=0.5056\textwidth]{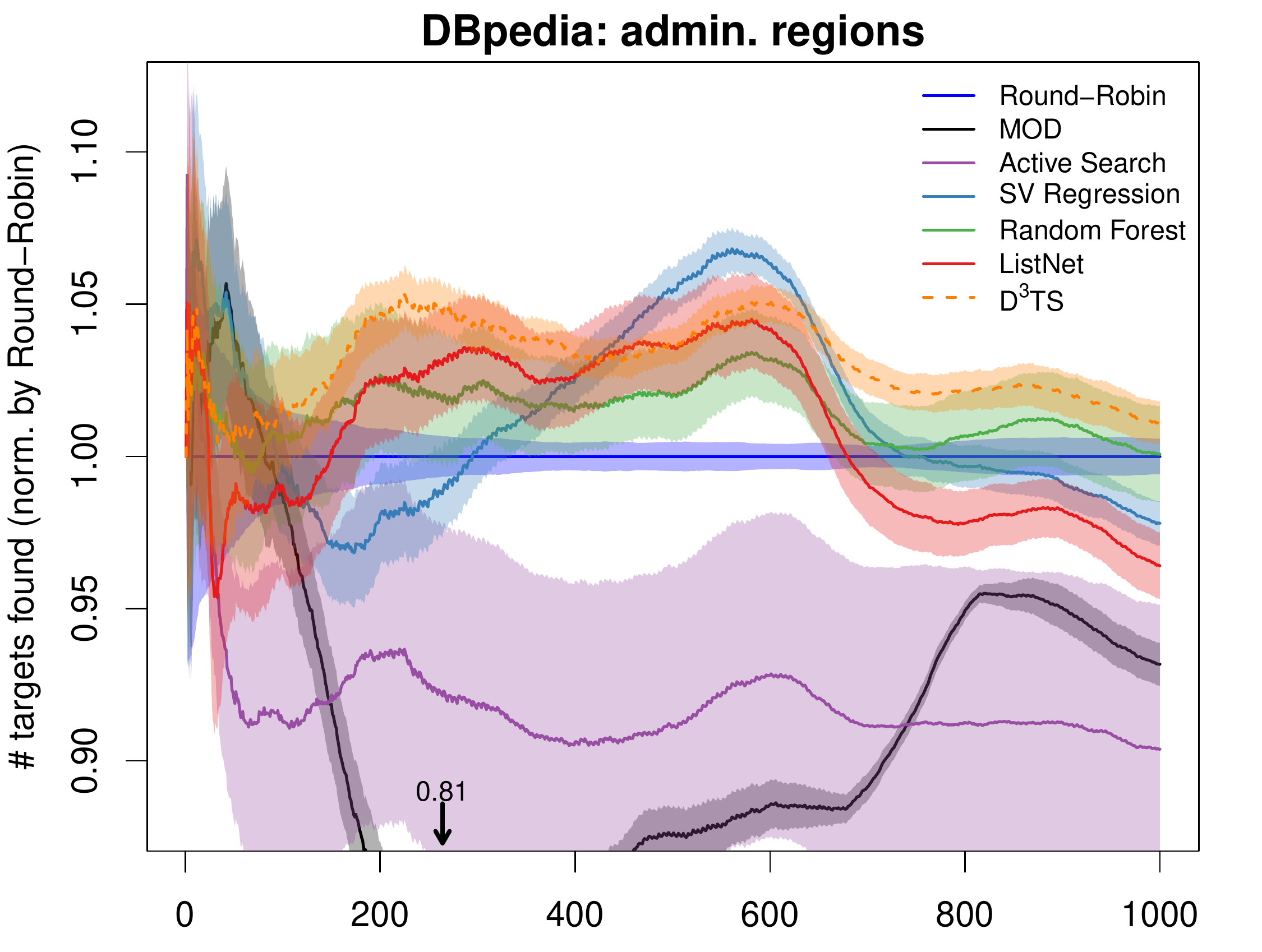}
     \label{fig:dbpedia_line}}
    \subfloat{\includegraphics[width=0.4944\textwidth]{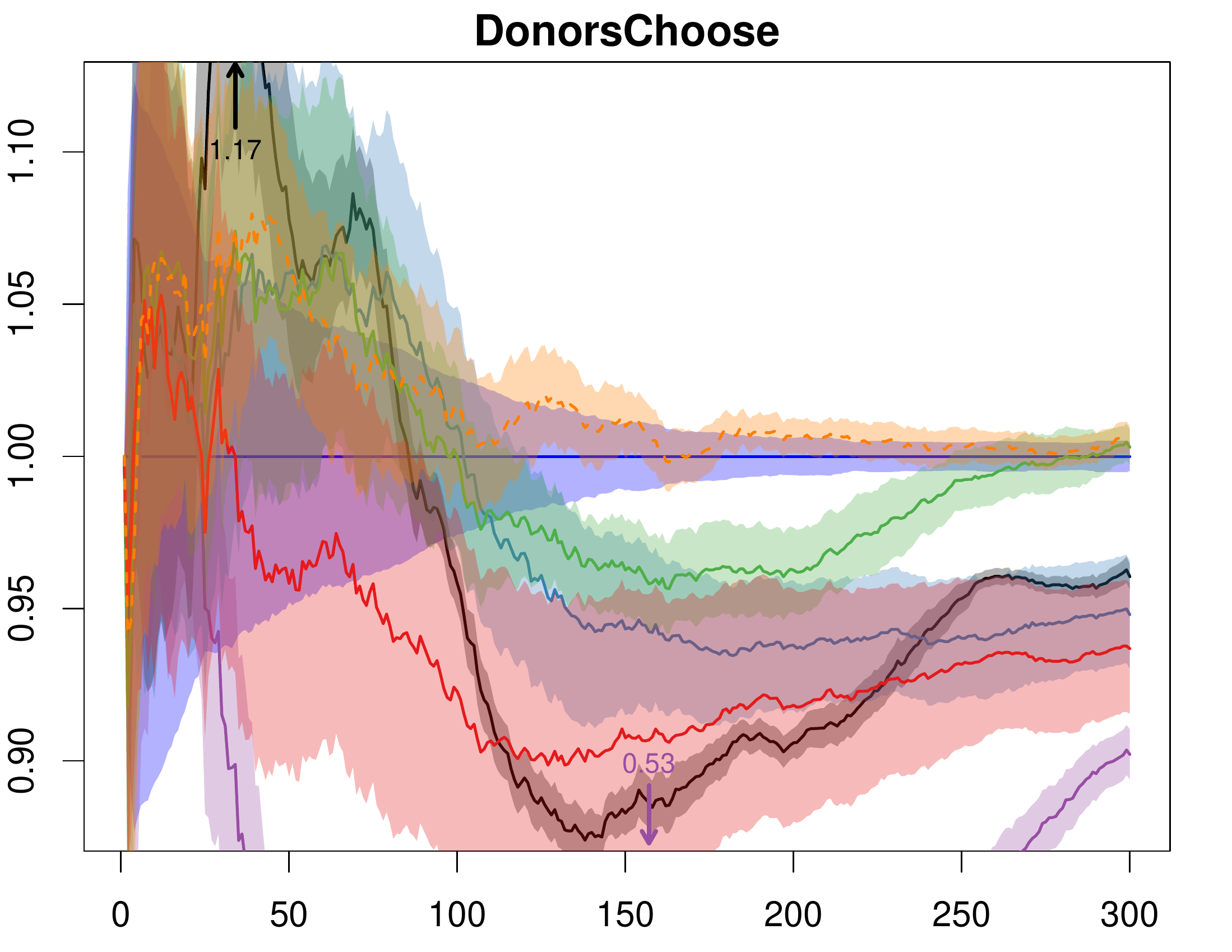}
    \label{fig:donors_line}}
    \\\vspace{-0.5cm}
         \subfloat{\includegraphics[width=0.5056\textwidth]{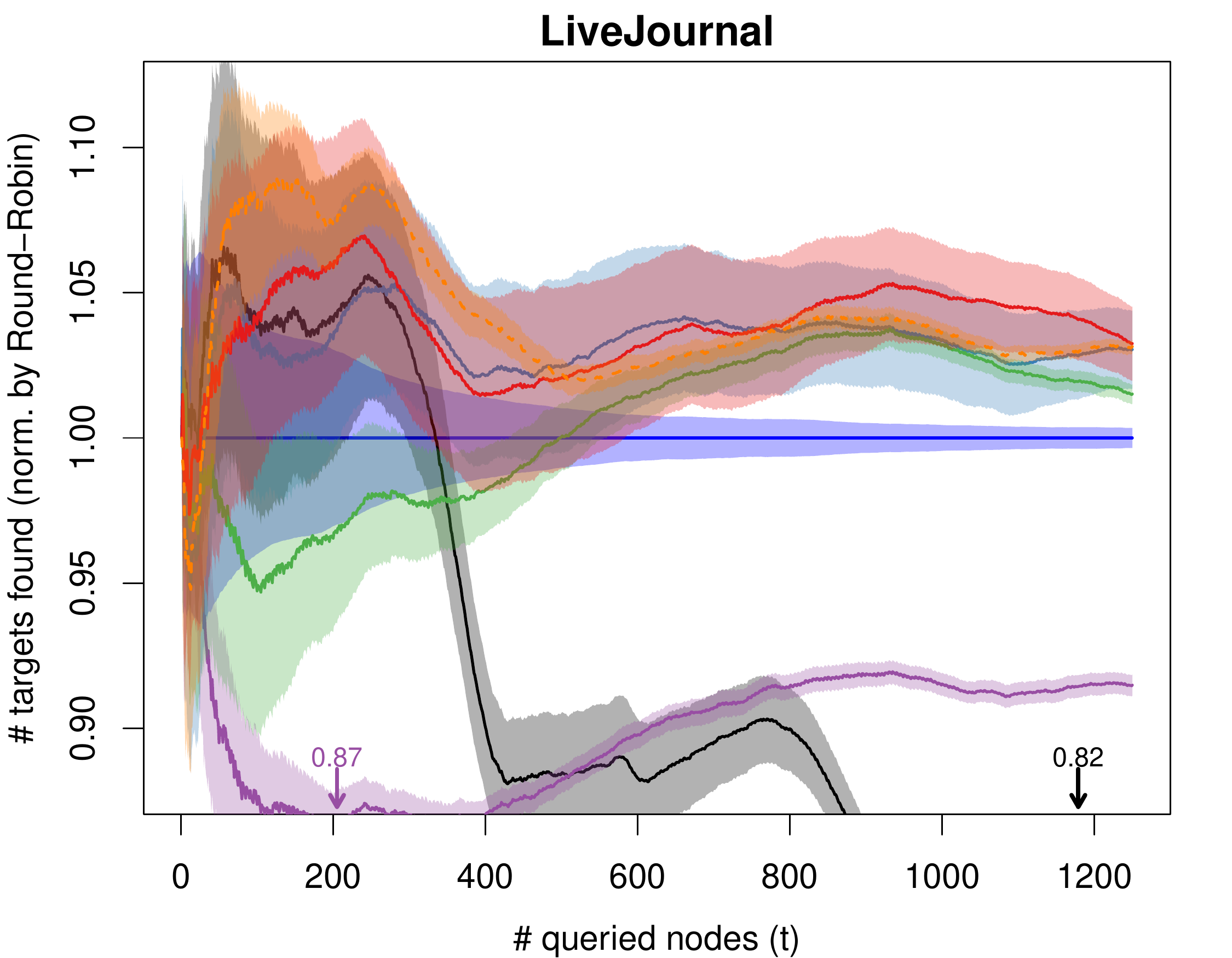}
     \label{fig:lj_line}}
         \subfloat{\includegraphics[width=0.4944\textwidth]{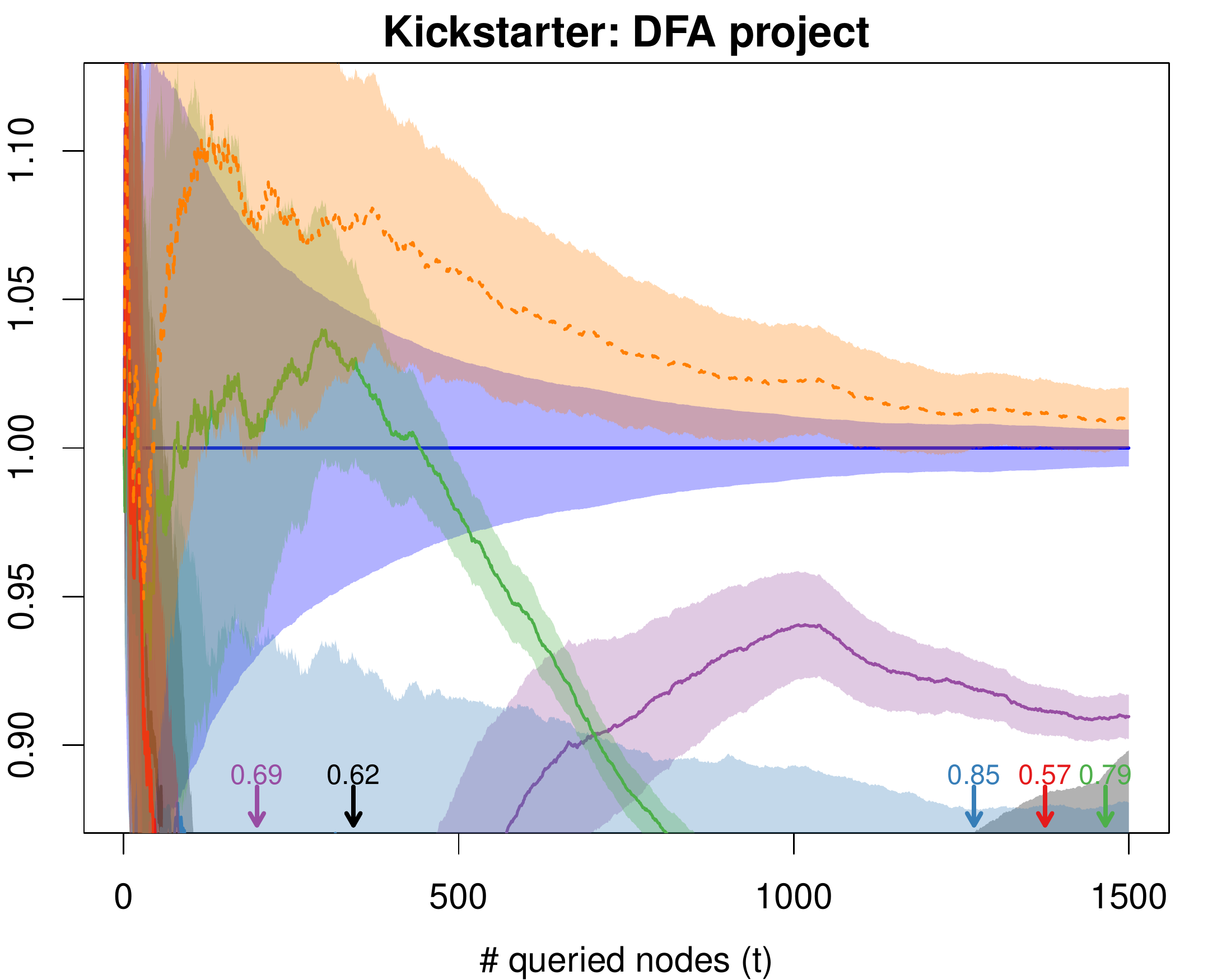}
     \label{fig:kickstarter_line}}
 
    %\subfloat{\includegraphics[width=0.25\textwidth]{images/single_models/donors_total_bars.pdf}
    %\label{fig:donors_all}}    
\caption{Average number of targets found by Round-Robin (RR), \ALGO and five standalone classifiers over 80 runs.
Shaded areas represent 95\% confidence intervals. Arrows indicate minimum values for corresponding colors' classifiers, when off-the-chart.
Standalone classifiers are often outperformed by RR. \ALGO improves upon RR.}
\label{fig:all}
\end{figure*}

We now describe the results for each dataset in detail, except for CiteSeer, which was discussed in the
introduction. Figure~\ref{fig:all} contrasts the average number of targets found by RR
and \ALGO against those found by standalone classifiers, scaled by RR's performance.
We include results for five out of nine classifiers (the same ones used in $\mathcal{M}$) to avoid clutter.

On DBpedia, LiveJournal, DonorsChoose and Kickstarter, even RR was able to outperform the
existing methods,
except for the initial steps (where absolute differences are small anyway). Moreover, on the first two datasets,
base learners outperformed existing methods. However, as shown in DonorsChoose and Kickstarter plots, a data-driven
classifier by itself does not guarantee good performance.

\begin{figure*}
\centering
\includegraphics[width=0.6\textwidth]{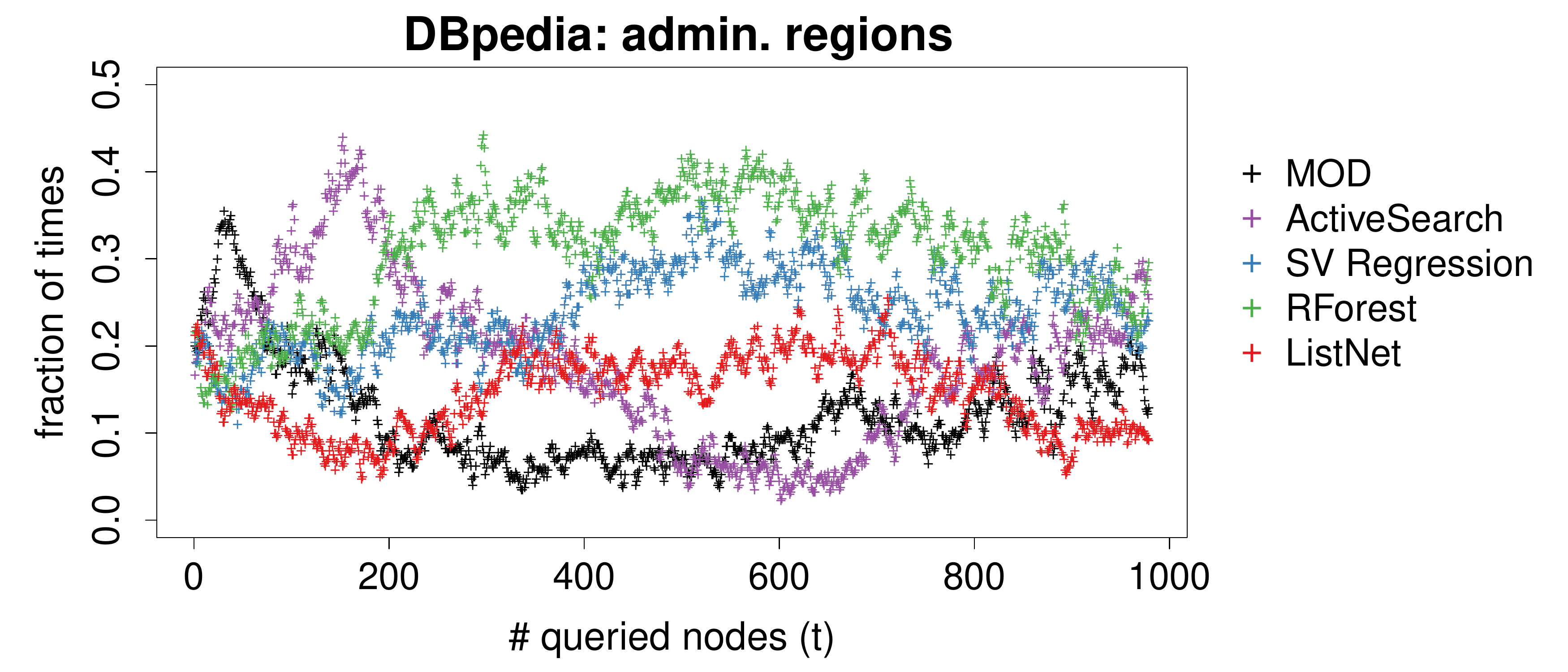}
\caption{\ALGO: fraction of runs in which each classifier was used in step $t$ (smoothed over five steps).}
\label{fig:convergence}
\end{figure*}

On most datasets \ALGO matches or exceeds the performance
of the best standalone classifier. In particular, on Kickstarter, both RR and \ALGO find significantly more target
nodes than standalone classifiers. While RR can leverage diversity from using multiple classifiers to avoid the tunnel vision effect,
\ALGO goes beyond and intelligently decides which classifier to use without harming diversity. To illustrate this,
we look at the fraction of times \ALGO used a given classifier at turn $t$ in 80 runs. Figure~\ref{fig:convergence}
shows this time series for DBpedia. From the small fraction of uses, we find that MOD performs
poorly not only on its own, but also when used under \ALGO. Fortunately, \ALGO can learn classifiers' relative performances and adjust
accordingly. 
%Despite exhibiting the lowest average performance, Active Search
%performs well in some simulation runs, which explains its large confidence interval in Fig.~\ref{fig:all}
%and its large fraction of uses seen towards the end in Fig.~\ref{fig:convergence}.

\begin{figure*}
  \center
        \subfloat{\includegraphics[width=0.34\textwidth]{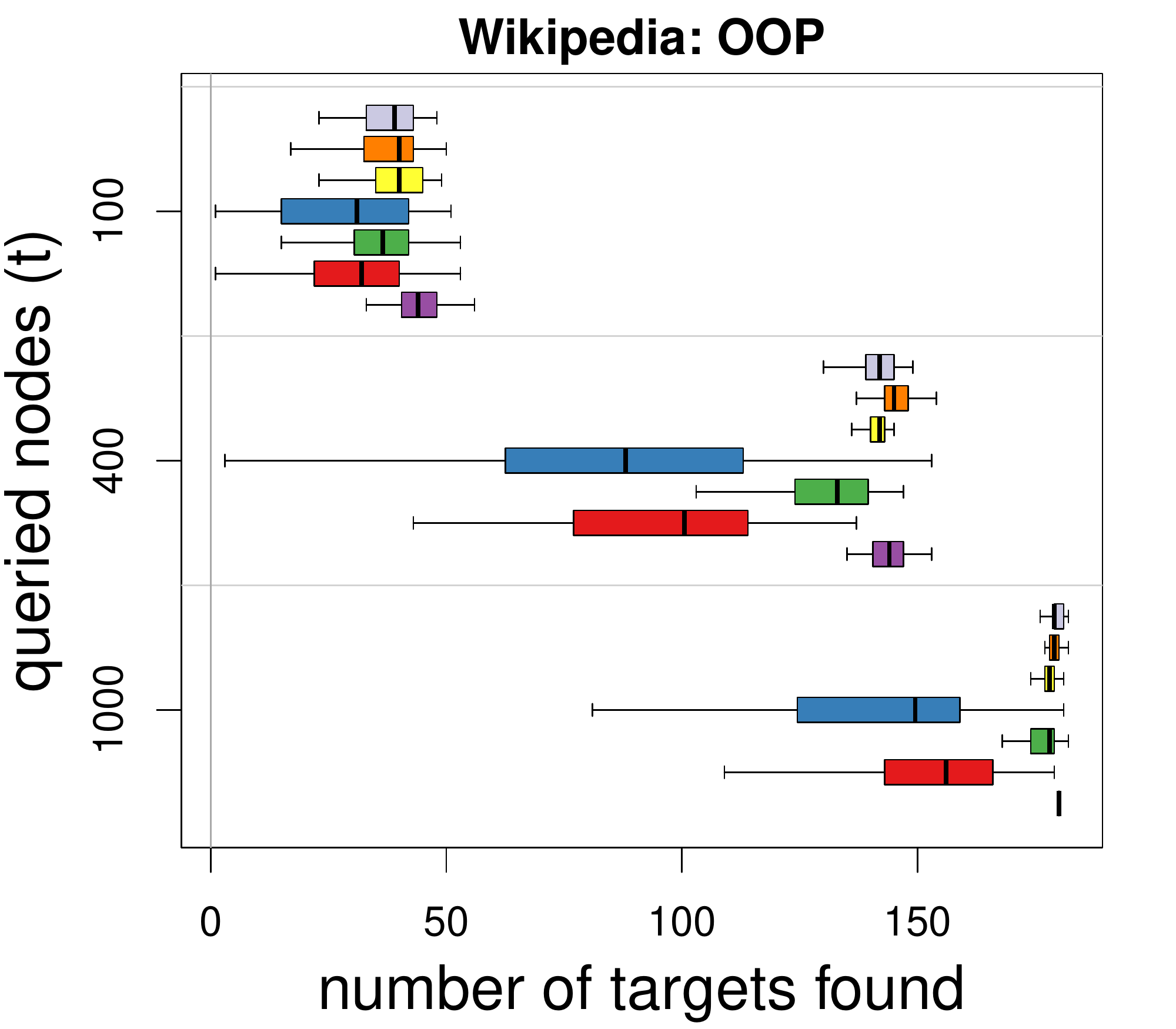}
     \label{fig:wikipedia_boxplot}}
    \subfloat{\includegraphics[width=0.33\textwidth]{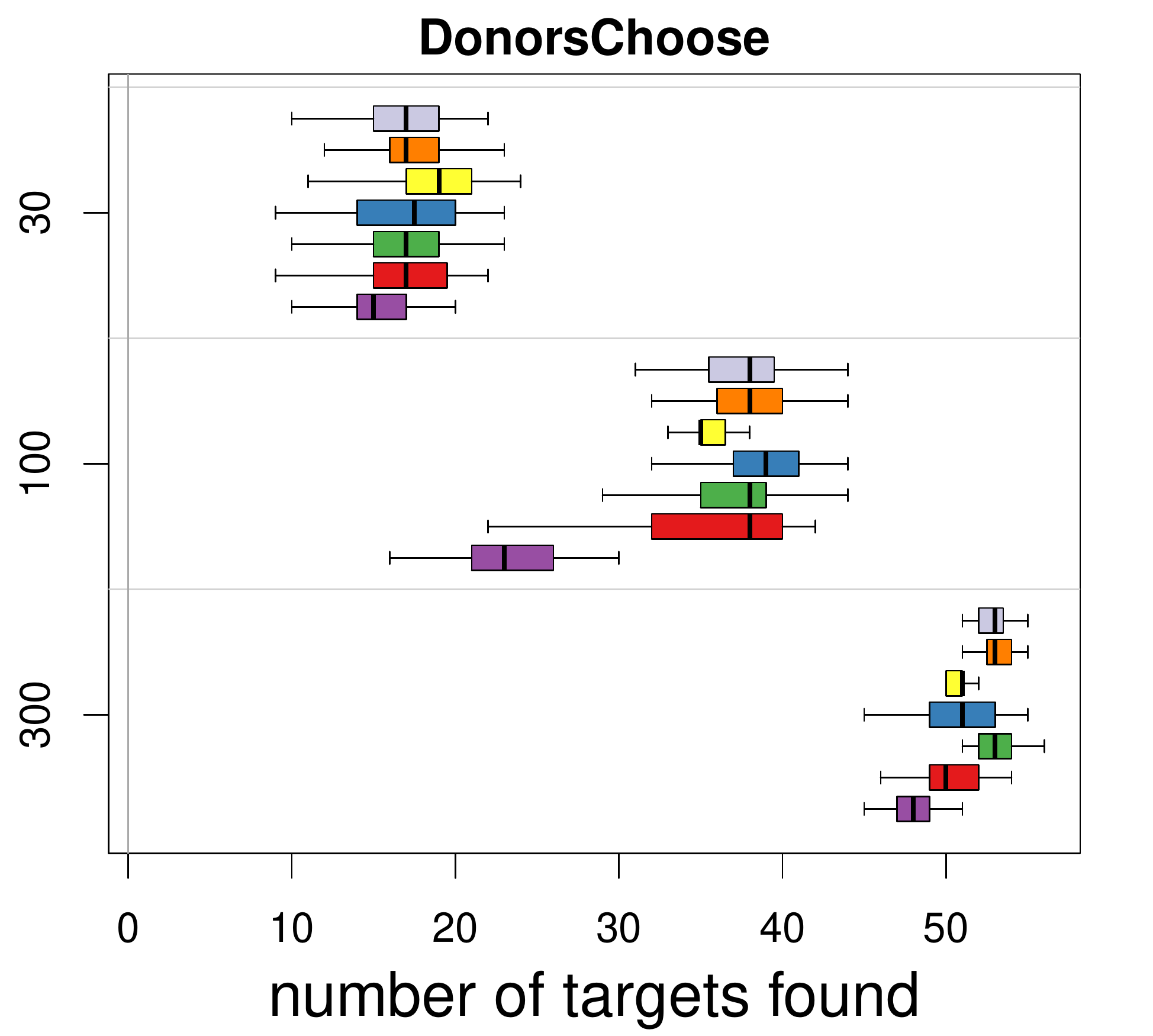}
    \label{fig:lj_boxplot}}
    \subfloat{\includegraphics[width=0.33\textwidth]{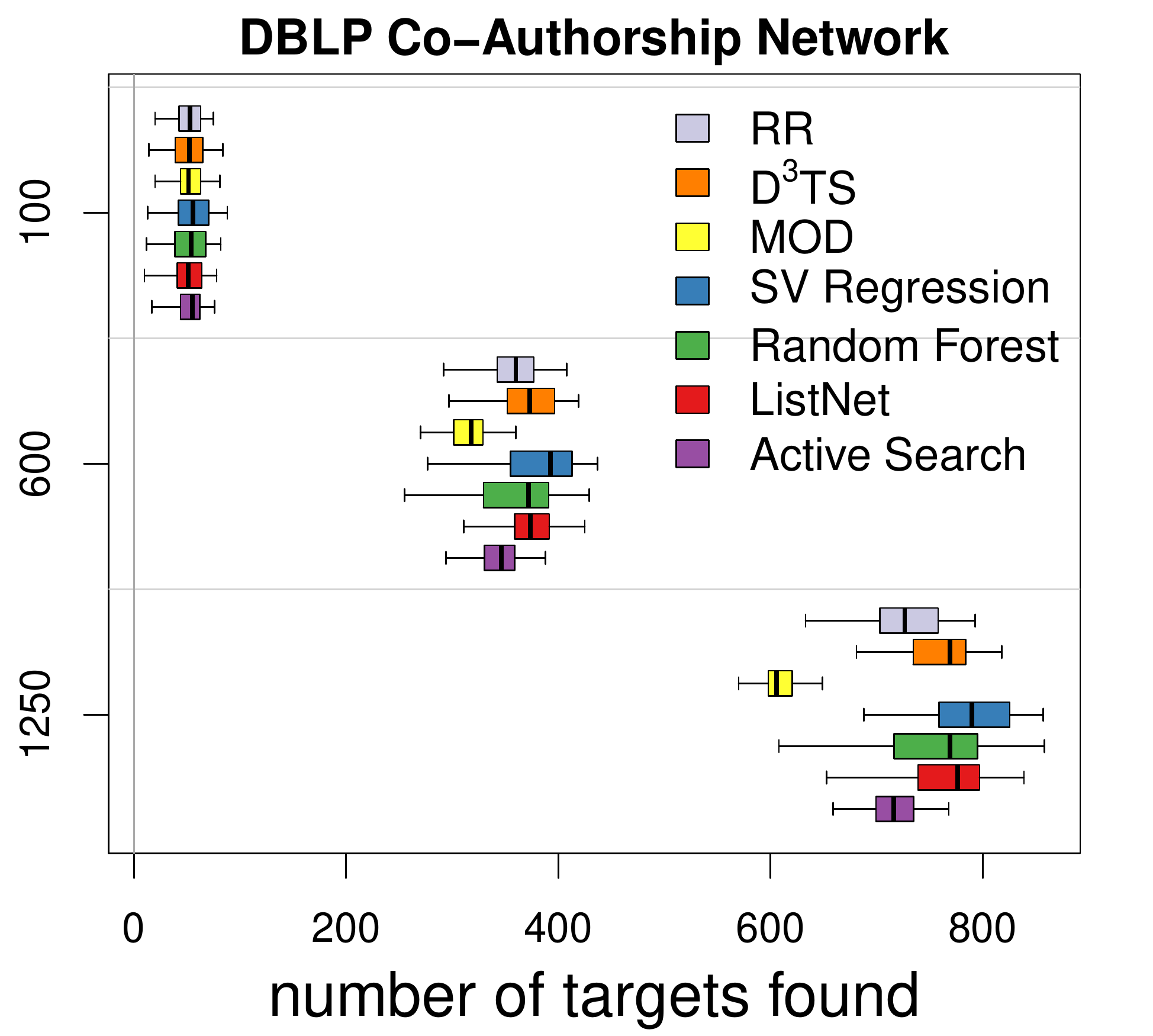}
     \label{fig:dblp_boxplot}}
\caption{RR and \ALGO can perform well even when including classifiers that perform poorly as standalone.}
\label{fig:boxplots}
\end{figure*}

A closer look at the distribution of the number of targets found by each method highlights an important
advantage of leveraging diversity. Figure~\ref{fig:boxplots} shows boxplots
of RR and \ALGO performance in each dataset, for several points in time.\footnote{The box extremes in our boxplots indicate lower and
upper quartiles of a given empirical distribution; its median in marked in between them. Whiskers indicate
minimum and maximum values.} On Wikipedia, DonorsChoose and Kickstarter,
although some of the classifiers used by RR and \ALGO yield poor results on their own,
RR and \ALGO still attain large mean and low variance.
\ALGO was only outperformed by a standalone classifier on DBLP (statistically significant).
Because DBLP has the largest number of target nodes in the border set (on average) over all datasets,
classifiers are less likely to be penalized by the tunnel vision effect on DBLP.

\fm{In Appendix~\ref{app:results} we provide complementary results from ten additional datasets derived from the same data. Once again these
results attest for the robustness of the proposed method.}

%Figures~\ref{fig:recall}(a,b) show...
%\begin{figure*}
%  \center
%    \subfloat[Line plot; make fonts bigger]{\includegraphics[height=0.6\textwidth,width=0.4\textwidth]{images/donors_recall_sorted_per_model}
%     \label{fig:recall-lines}}
%        \subfloat[Graph heatmap]{\includegraphics[width=0.6\textwidth]{images/donors_occupancy}
%     \label{fig:recall-heatmap}}
%\label{fig:recall}
%\caption{META: use both figures? use earlier snapshot, say 50\% found; replace RR by DTS.}
%\end{figure*}    

\subsection*{Classifier combinations}

We also conducted an exhaustive set of simulations where we consider all 31 combinations
of these five classifiers under \ALGO. We restrict this analysis
to a set of networks $\bm{\cD}$ composed of the five smaller datasets.
Suppose we had an oracle that could tell which
combination of classifiers performs best on a dataset $\cD \in \bm{\cD}$.
We can then define the (normalized) regret of a classifier set $\cM$ on
$\cD$ as
\[
  R(\cM,\cD) = 1 - \frac{ N_+(\cM,\cD) } {\max_{\cM^\prime} N_+(\cM^\prime,\cD)}
\]
where $N_+(\cM,\cD)$ is the number of target nodes found by $\cM$ on $\cD$.
If we define the optimal combination to be the one that minimizes the maximum regret,
i.e., $\cM^\star = \arg\min_{\cM} \max_{\cD \in \bm{\cD}} R(\cM,\cD)$, then
$\cM^\star$ indeed includes all five classifiers (maximum regret is 2.8\%).
Otherwise, if we define the optimal
combination $\cM^\dagger$ to be the one that minimizes the average regret,
i.e., $\cM^\dagger = \arg\min_{\cM} \sum_{\cD \in \bm{\cD}} R(\cM,\cD)/|\bm{\cD}|,$ then
$\cM^\dagger$ is the combination composed of MOD, Active Search, SVR and Random Forest
(average regret is 0.9\%). We note, however, that the performance obtained by
combination $\cM^\star$ on each dataset is at most 0.7\% smaller than that
obtained by $\cM^\dagger$ (in the case of CiteSeer). 
\fm{Moreover, we observed that combining two classifiers improves results in about 84\% of the cases
w.r.t.\ the cases where either classifier is used in isolation}. This
attests to the robustness of using \ALGO as the classifier selection policy.

\subsection*{Running time}

\begin{table*}[t]
\centering
\begin{tabular}{lrrrrrrr}
  \hline
   \multirow{2}{*}{{\bf Methods}} & \multicolumn{7}{c}{{\bf Datasets}}\\
   \cline{2-8}
                     & \multicolumn{1}{c}{{\bf CS}} & \multicolumn{1}{c}{{\bf DBP}}& \multicolumn{1}{c}{{\bf WK}} & \multicolumn{1}{c}{{\bf DC}} & \multicolumn{1}{c}{{\bf KS}} & \multicolumn{1}{c}{{\bf DBL}} & \multicolumn{1}{c}{{\bf LJ}} \\ 
\hline
  MOD                  & 0.06 & 0.08 & 0.14 & 0.29 & 3.55 & 0.33 & 0.46\\
  Active Search    & 0.05 & 0.11 & 0.17 & 0.37 & 1.71 & 0.30 & 0.45\\ 
  SV Regression  & 0.37 & 0.80 & 1.26 & 5.88 & 9.35 & 6.57 & 8.19 \\ 
  Random Forest &  2.54 & 4.27 & 6.75 & 16.75 & 43.80 & 20.96 & 21.06 \\ 
  ListNet              & 0.35 & 0.31 & 1.76 & 2.13 & 8.42 & 22.65 & 21.85 \\
Round-Robin      &  0.18 & 0.14 & 0.31 & 0.34 & 2.49 & 11.13 & 10.92 \\ 
  \ALGO              &  0.13 & 0.14 & 0.28 & 0.27 &  2.77 & 13.41 & 13.28 \\
  \end{tabular}
  \caption{Average wall-clock time to find a target (in sec.). \ALGO benefits from more sophisticated classifiers while only incurring the computational cost for the steps in which they are used.}
  \label{tab:times}
  \end{table*}

Table~\ref{tab:times} shows the average wall-clock time to find a target based on 80 single-threaded runs on an Intel Xeon E5-2660@2.60GHz processor for MOD, Active Search, SVR, Random Forest, ListNet, RR and \ALGO. On all datasets except DBLP and LiveJournal, Random Forest is based on conditional
inference trees (from R package \verb|party|), which are recommended when different types of features (e.g., discrete, continuous) are present~\cite{hothorn2006unbiased}. On the other two datasets, Random Forest is based on classical decision trees (from R package \verb|randomForest|), due to the large scale of these datasets. In both cases the average number of targets found was similar, but conditional inference trees tend to yield smaller variances.

Among standalone classifiers, MOD and Active Search were the fastest, followed
by ListNet and SVR. We emphasize that MOD and Active Search require no fitting, which is the most expensive step for a base learner. In spite of their good performance at finding target nodes on DBLP and LiveJournal, Random Forest and ListNet take much longer to fit than other classifiers on datasets with a relatively large number of features, thus exhibiting the longest average time between successful queries.

One of the advantages of \ALGO is that it can benefit from
more sophisticated classifiers while only incurring the
computational cost for the steps in which they are used.
\ALGO exhibits smaller ratios than
Round-Robin, except on datasets where \ALGO tends
to use Random Forest or ListNet more often than Round-Robin does.
Note that \ALGO running time is determined by the classifiers it uses and their implementations. Replacing methods used in this paper by online counterparts can lead to significant reductions in running time. In particular, Random Forest -- which has the largest running time -- can, in principle, be replaced by
online random forests when bounds on feature values are known in advance.\footnote{We attempted to replace Random Forests by Mondrian Forests \cite{lakshminarayanan2014mondrian}, but the only publicly available implementation is not optimized enough to be used in our application.}

\subsection{Dealing with Disconnected Seeds}

%  META: here we go back to the discussion of single models and talk about
%  whether it makes sense to use one model for each region without "pooling".
 %  Show that even with partial pooling, the results were not better than the case
%  of one single model for all regions. Argue that there is no reason why we should
%  not use Round Robin or MAB algorithms here.

In the previous simulations, the search starts from a single seed (starting node). When more than one seed is available, the search process may end up exploring various regions of the graph at the same time. In this approach, the question arises as to how to adequately model the observations in these regions. 
%It is possible that some regions posses very different target availability and induce difference correlations between the observed graph and border node labels.
In some cases,
it is better to fit classifiers to specific regions of the network where they operate (i.e., using observations collected only from that region), while fitting all classifiers to all observations is probably best all regions are very similar to each other. One can also consider hierarchical models, which model each region separately but allow some information 
sharing. %estimated over the entire observed network.

In this section, we consider standalone classifiers and compare their performance in two extreme approaches: using a single classifier and starting from $S$ seeds (thus modeling all $S$ regions together), or using $S$ models, each initially associated with a single seed (each simulation run uses the same $S$ seeds in either approach to reduce variance). In particular, we use the EWLS regression model.

%The single-model scenario bears no difference to the standalone models previously descibred.
In the multiple classifier approach, the classifier associated with each region is used to rank its corresponding border set at each time $t$. A single node to be queried must then be selected among all border nodes. We select the node with the highest estimated payoff across all rankings, and the model responsible for this estimation is then updated with the new observation. 

We compare the search performance under these two approaches, for $S=2,\ldots,6$. On the datasets with larger number of attributes, we found that either there is no significant difference between the average payoffs (Donors, CiteSeer) or the single classifier approach yields better performance (Wikipedia), at the $95\%$ confidence level.
On the other hand, on datasets with a small number of attributes, some improvement is obtained when using multiple classifiers, each with its own model. For instance, on DBpedia, which has only $5$ attributes, the average number of targets found increases from $523.9$ to $562.5$ \fm{at t=1000}, for $S=3$.

When \ALGO is used in place of standalone classifiers, our recommendation is to fit base learners to region-specific observations in the case of datasets with few attributes, and fit to the entire training set in the case of datasets with many attributes. However, if new seeds are included during the search (i.e., $S$ increases over time), it is likely beneficial to fit the initial classifiers corresponding to the new regions using observations from other regions as priors, even if the number of attributes is large. We leave this investigation for future work.

% Activate the following line by filling in the right side. If for example the name of the root file is Main.tex, write
% "...root = Main.tex" if the chapter file is in the same directory, and "...root = ../Main.tex" if the chapter is in a subdirectory.
 
%!TEX root =  main.tex

\section{Related work}\label{sec:related}

The closest work to ours is on active search. The goal of active search is to
uncover as many nodes of a target class as possible in a network where the
topology is known~\cite{Garnett:2011wt,garnett2012bayesian,Wang2013,Ma:2015ut}.
Like \probname, active search considers situations
%such as fraud detection or the investigative analysis of potentially criminal social networks,
where only members of a target class (e.g., malicious class) are sought. Since obtaining labels is associated with a cost (time or money), it is paramount to avoid spending resources on nodes that are unlikely to be targets.
Unlike our problem, active search assumes the network topology is known and that any node can be queried at any time.

In \cite{Pfeiffer:2013cikm} a problem similar to \probname is investigated and a
learning-based method called Active Exploration (AE) is proposed. Unlike
in \probname, border nodes attributes are assumed to be observable.
Since node attributes often carry considerable information about the node's label,
AE is not directly comparable with other \probname methods. Our solution
differs from AE in that it leverages heuristics in addition to
base learners and is applicable to a wider range of applications.
Similarly to \probname, active learning is an interactive framework for deciding
what data points to collect in order to train a classifier or a regression model. Unlike active search, 
%Active learning is an interactive framework conceived to explicitly model the process of obtaining labels for unlabeled data. However,
(i) its main objective is to improve the generalization performance of a model
with as few label queries as possible,
and (ii) the set of unlabeled points does not grow based on the collected points.
A slew of active learning techniques have been proposed for non-relational data settings,
including some tailored for logistic regression
\cite{schein2007}, for dealing with streamed data
\cite{Attenberg:2011hw} and for the case of extreme class imbalance \cite{Attenberg:2010vk}.
Although the retrieval of target nodes can benefit from an accurate model, it is unlikely
that active learning heuristics (e.g., uncertainty sampling~\cite{settles2010active}) for training a single classifier
can be used for \probname without sacrificing performance. However, it may be possible to adapt active learning techniques
proposed for training classifier ensembles (e.g., query by committee~\cite{seung1992query}) in such a way that, at the same time we
collect points on which many classifiers disagree, we ensure that promising candidates among border nodes are queried before
the sampling budget is exhausted.

Despite these differences, there is an interesting parallel between \probname
with many models and a body of research on active learning with a set of active
learners (or heuristics). Both problems can be cast as MABs, where border
nodes are analogous to unlabeled data points. In active learning, a reward
is indirectly related to the collected point: it is computed as some
proxy for or estimate of the model's performance on a test set,
when fit to all points collected up to a given step. In
contrast, rewards in \probname are simply the node labels. Like \probname,
active learning can either map heuristics directly as arms
\cite{Baram:2004uc} or map heuristics as experts that give recommendations on
how to choose the unlabeled points \cite{Hsu:2015wm}. In both cases it has
been observed that combining heuristics may often outperform the single best
heuristic. While these works apply algorithms for adversarial bandits to active
learning, we find that Dynamic Thompson Sampling for stochastic bandits with
non-stationary rewards seem to exploit better the fact that arms rewards are
slowly changing in \probname.

Last,
another variant of active learning considers the task of learning an ensemble of models \cite{Ali:2014vl}
or finding a low risk hypothesis $h \in \mathcal{H}$ \cite{Ganti:2012tq,Ganti:2013wc} while
labeling as few points as possible. Since the labeled points are biased by
the collection process, estimating the models' generalization performances requires either
building an uniformly random validation set, or sampling probabilistically at every step
and then using importance weighted estimates. In \probname, however, the models relative
performances can be directly measured from the queried nodes payoffs. Moreover,
building a random validation set is bound to degrade performance in
scenarios where target nodes are scarce.

\section{Discussion} \label{sec:discussion}

\fm{In this section, we discuss the technical challenges in accounting for the
future impact of a query and contrast the proposed solution with classical ensemble learning.}

\subsection{Accounting for the future impact of querying a node}\label{sec:impact}

Active search assigns
a score to each potential border node $v$ that consists of a sum of two terms \cite[eq.\ (2)]{Wang2013}:
the expected value of $v$'s label and sum of the expected changes in the labels
of all other nodes multiplied by a discount factor $\alpha \ll 1$.
The discounted term tries to account for the impact of querying node $v$,
going one step beyond the greedy solution.
In \probname, however, the observed graph is limited to the set of queried nodes and their
neighbors, i.e.\ we cannot compute the impact of choosing a node beyond the border set.
Even if we could observe the entire graph, accounting for the future impact
of querying a node would
require us to fit one statistical learning model to each border node
and predict all the remaining labels at each step,
which is too expensive even for a single online model.

\subsection{Using classifier ensembles in \probname}

Ensemble methods generate a set of models in order to combine their predictions, possibly using weights.
These methods perform very well in many classification problems
and can be applied to \probname problems too. Note that although \ALGO uses multiple statistical models, it cannot be considered a classifier ensemble, since
only one classifier is used for prediction at each step.

%Bagging techniques, construct subsets of training data by sampling ...

\begin{table*}[t]
\centering
\begin{tabular}{lrrrrr}
  \hline
   \multirow{3}{*}{{\bf Methods}} & \multicolumn{5}{c}{{\bf Datasets} (budget $T$)}\\
   \cline{2-6}
                     & \multicolumn{1}{c}{{\bf CS}} & \multicolumn{1}{c}{{\bf DBP}}& \multicolumn{1}{c}{{\bf WK}} & \multicolumn{1}{c}{{\bf DC}} & \multicolumn{1}{c}{{\bf KS}} \\ 
    & (1500)    & (700) &  (400) & (100)    & (700)  \\
\hline
  Bagging &  745.6 & 445.6 & 99.1 & 34.7 & 223.1  \\ 
   AdaBoost & 751.5 & 443.5 & 98.0 & 34.5 & 218.4 \\  
  \ALGO & 851.2 & 464.0 & 144.7 & 37.9  & 247.6  \\
  Bootstrap + Decision Tree &  754.5 & 293.4 & 95.2 & 27.2 & 155.7  \\ 
  \hline
\end{tabular}
\caption{Average number of targets found by each method after $T$ queries based on 80 runs.}
\label{tab:ensemble}
\end{table*}

We simulate two popular ensemble methods -- Bagging and AdaBoost -- on five datasets
(DBLP and LiveJournal were not included due to the prohibitive execution time).
For Bagging, we varied the number of trees in $\{5, 10, 100\}$, minimum number of observations to split a node in $\{5, 10\}$
and maximum tree depth in| $\{1, 5, 10\}$. For Boosting, we set the maximum tree depth to 1 and varied the number of trees in $\{100, 200\}$.
Table~\ref{tab:ensemble} displays the results associated with the configurations that obtained the best overall results -- Bagging(\verb|ntree|=100, \verb|minsplit|=10, \verb|maxdepth|=5) and Boosting(\verb|maxdepth|=1, \verb|ntree|=100) --
along with the results obtained by \ALGO. We find that \ALGO consistently outperforms these ensemble methods.
We conjecture that ensembles are only slightly less susceptible to the tunnel
vision effect than standalone models, as combining predictions tends to decrease border set and training set diversity.

What if we do not combine their predictions? In other words, what if we generate a decision tree from bootstrap sampling at each step
and use that to make predictions? We simulated the performance of this mechanism, varying the minimum number of observations to split a node in $\{5, 10\}$
and maximum tree depth in $\{5, 10\}$. However, this approach did not perform as well as \ALGO (or even RR). We report in Table~\ref{tab:ensemble} the parameter configuration that achieved the best overall results, (\verb|minsplit|=10, \verb|maxdepth|=10),  under ``Bootstrap + Decision Tree''. The poor performance of this approach can be explained by the fact that predictions made from a single tree are not very accurate.
By making predictions with a single tree, we lose the generalization benefits that come from classifier ensembles.
%Yet, ``Bagging w/o combining'' outperformed Bagging on the Wikipedia dataset.
%Similar to classic ensembles, the accuracy of individual models is important
%for this mechanism to perform well.
%On the other hand, the diversity sought in \probname is different than the one sought in ensemble learning,
%as we explain next.

%In what follows, we contrast the diversity achieved by ensembles with that desirable in \probname problems.

\subsection{Contrasting diversity in ensembles and diversity in \probname}

%Ensemble methods generate a set of models in order to combine their predictions, possibly using weights.
Diversity is known to be a desirable characteristic in ensemble methods \cite{kuncheva2003elusive,tang2006analysis,xie2016diversity}. The intuition is
that if one can combine accurate models that make uncorrelated mistakes, the overall
accuracy will be higher than those of the individual models.
There are two main classes of techniques for generating diverse ensembles \cite{stapenhurst2012diversity}:
(i) {\em overproduce and select}, where a large set of base learners is generated, among which a subset is selected to maximize a given measure of diversity, (ii) {\em building ensembles}, where the diversity measure is directly used to drive the ensemble creation.
In the ensemble literature there are several metrics proposed for quantifying diversity, all of which can be computed from the predictions made by different models.
Many of these metrics are shown to have positive correlation with the overall accuracy of the ensemble.

In \probname, the relationship between correlations in models' mistakes and overall performance is more indirect.
For a single query, whether mistakes made by different models are uncorrelated or not is immaterial, since we use
only one model to decide which node to query at each step. On the other hand, every query choice impacts future steps.
Therefore, differences in models' predictions dictate the levels of border set and training set diversity that will be
achieved over time. This is in sharp contrast with the static notion of diversity referred in the ensemble literature.
A deeper characterization of the sets of models
that can achieve the type of diversity that leads to good performance in \probname is left as future work.

%\subsection{Obtaining different classifiers by fitting to subsets of the training data (bagging)}
%
%We show in Table~\ref{tab:best} that classifier ensembles obtained through bagging do not perform as well as RR or \ALGO. However,
%the reader may wonder what happens if we use such ensembles without combining classifiers predictions, that is,
%selecting one of the classifiers to make predictions at each step. In other words: what if we generate a set of classifiers $\cM$
%from the same family by fitting them to different subsets of the training data (bagging)?
%We simulated the performance of fitting an ensemble of decision trees using bagging and then randomly selecting one of the trees to predict border node labels at each step, but this approach did not perform as well as RR. These results are shown in Figure~XXX.}

\section{Conclusions} \label{sec:conclusions}

This paper introduced \probname, where the goal is to find the
largest number of target nodes given a fixed budget and subject to a partial --
but evolving -- understanding of the network.
\fm{The key distinctions of selective harvesting w.r.t.\ related problems are that (i) the network is not fully observed and/or (ii) a model must be learned during the search. These distinctions combined make the problem much harder than the related problems.} 
%In particular, we showed that existing solutions that can be applied or adapted to selective harvesting suffer from the tunnel vision effect.
We discussed existing methods
that can be adapted to \probname and an alternative approach based on statistical
models. However, we showed that the tunnel vision effect incurred by the nature
of the \probname task severely impacts the performance of a classifier
trained on these conditions. We show that using multiple classifiers is helpful in mitigating the
tunnel vision effect. In particular, simulation results showed that methods used in isolation often perform worse than when combined through a
round-robin scheme.
We raised two hypothesis to explain this observation, which were investigated to
show that classifier diversity -- i.e., switching among classifiers at each querying step
-- is important for collecting a larger set of target nodes in \probname.
Classifier diversity increases the diversity of the
training set while broadening the choices of nodes that can be queried in the
future.
Based on these observations we proposed \ALGO, a method based on multi-armed
bandits and classifier diversity, able to account for what we named the exploration,
exploitation and diversification trade-off. \fm{\ALGO differs from traditional ensembles,
in which it does not combine predictions from different models at a given step. \ALGO
also differs from traditional MABs, in which the goal is not to converge to a single arm.}
\ALGO outperforms all competing
methods on five out of seven real network datasets and exhibited comparable performance
on the others. While we evaluated \ALGO's performance when used with five specific classifiers (MOD, Active Search,
Support Vector Regression, Random Forest and ListNet),
the proposed method is flexible and can be used with any set of classifiers
(not shown here, replacing SVR with Logistic Regression yielded similar results).
Moreover, we showed that combining two classifiers through \ALGO improves results in about 84\% of the cases
w.r.t.\ the cases where either classifier is used in isolation.

%This suggests that, at the same time that \ALGO benefits from 
%the inclusion of good-performing classifiers, it is robust to the inclusion of poor-performing ones.
%ACKNOWLEDGMENTS are optional
\begin{acknowledgements}
This work was sponsored by the ARO under MURI W911NF-12-1-0385, the U.S. Army Research Laboratory under Cooperative
Agreement W911NF-09-2-0053, the CNPq, National Council for
Scientific and Technological Development - Brazil, FAPEMIG, NSF under SES-1230081, including support from the National Agricultural Statistics Service. The views and
conclusions contained in this document are those of the author and should not be interpreted as
representing the official policies, either expressed or implied of the ARL or the U.S.\ Government.
The U.S.\ Government is authorized to reproduce and distribute reprints for Government
purposes notwithstanding any copyright notation hereon. The authors thank Xuezhi Wang
and Roman Garnett
for kindly providing code and datasets used in \cite{Wang2013}.
\end{acknowledgements}

\bibliography{recruit} 
\bibliographystyle{spbasic}

\appendix

\section{Complementary results}\label{app:results}

%what did we do
\fm{In Section~\ref{sec:results} we presented results obtained when defining the target populations either as in prior work
or as the largest subpopulation in the network. We extend these results by running simulations on ten additional datasets
derived by taking the two largest subpopulations as targets (other than the original targets) from CiteSeer, DBpedia, Wikipedia, DonorsChoose
and Kickstarter. These datasets are indicated by CS, DBP, WK, DC and KS, followed by 1 and 2, respectively. 
  Table~\ref{tab:complementary} shows performance results for five standalone models and for their combinations using Round-Robin and \ALGO.
  Except for DBP1 and WK1, \ALGO consistently figures among the two best performing methods.}

\begin{table*}[ht]
\centering
\begin{tabular}{rllllllllll}
\hline
  \multicolumn{1}{c}{{\bf Methods}} & \multicolumn{10}{c}{{\bf Datasets}}\\
   \cline{2-11}
%                     & \multicolumn{1}{c}{{\bf CS1}} & \multicolumn{1}{c}{{\bf CS2}} & \multicolumn{1}{c}{{\bf DBP1}} & \multicolumn{1}{c}{{\bf DBP2}} & \multicolumn{1}{c}{{\bf WK1}}& \multicolumn{1}{c}{{\bf WK2}} & \multicolumn{1}{c}{{\bf DC1}}& \multicolumn{1}{c}{{\bf DC2}} & \multicolumn{1}{c}{{\bf KS1}} & \multicolumn{1}{c}{{\bf KS2}} \\ 
 & CS1 & CS2 & DBP1 & DBP2 & WK1 & WK2 & DC1 & DC2 & KS1 & KS2 \\
  \hline
MOD  & 673 & 431 & {\bf 581} & 436 & {\bf  79} & {\bf 128} &  23 & {\bf  20} & 126 & 163 \\
  Active Search  & 666 & {\bf 568} & 550 & 403 & {\bf  79} & 124 &  15 &  10 & 115 & 213 \\
  SV Regression & 615 & 492 & 515 & 428 &  71 &  91 &  22 &  18 & 161 & 200 \\
  Random Forest  & 596 & 498 & 524 & 406 &  77 & 104 &  23 &  18 & {\bf 183} & {\bf 246} \\
  Round-Robin & {\bf 675} & 561 & {\bf 569} & {\bf 439} &  70 & 124 & {\bf  23} &  18 & 175 & 239 \\
  \ALGO  & {\bf 675} & {\bf 562} & 557 & {\bf 450} &  72 & {\bf 128} & {\bf  23} & {\bf  18} & {\bf 191} & {\bf 240} \\
   \hline
\end{tabular}
\caption{Simulation results on ten datasets derived from the original data attest. Best two methods on each dataset
are shown in bold. \ALGO performs consistently well.}
\label{tab:complementary}
\end{table*}
  
\section{Can we leverage diversity using a single classifier?} \label{app:random}

% why the single model cannot correct itself, or achieve diversity?
Intuitively, when a learning model is fitted to the nodes it chose to query, it tends to specialize in one region of the feature space and the search will consequently only explore similar parts of the graph, which can severely undermine its potential to find target nodes.

One potential way to mitigate this overspecialization would be to sample nodes probabilistically, as opposed to deterministically
querying the node with the highest score. Clearly, we should not query nodes uniformly at random {\em all the time}.
It turns out that querying nodes uniformly at random {\em periodically} does not help either, according to the following experiment.
We implemented an algorithm for \probname that samples at each step $t$, with probability $p$, an uniformly random node from $\mathcal{B}(t)$,
and with $1-p$, the best ranked node according to a support vector regression (SVR) model. Table~\ref{tab:random} shows the results for $p=2.5$, $5.0$, $10$, $15$ and $20\%$.
% latex table generated in R 3.2.2 by xtable 1.8-0 package
% Wed Feb 10 13:06:51 2016
\begin{table}[h]
\centering
\scriptsize
\begin{tabular}{ccccccc}

\hline
 0.0\% & 2.5\% & 5.0\% & 10\% & 15\% & 20\%  \\
\hline
 $760.5 \pm 52.1$ & $773.85 \pm 34.5$ & $768.0 \pm 32.3$ & $770.8 \pm 34.1$ & $753.0 \pm 59.8$ & $764.7 \pm 28.0$  \\
   \hline
\end{tabular}
\caption{Results for SVR w/ uniformly random queries on CiteSeer (at $t=1500$) averaged over 40 runs.
Top line shows probabilty of random query; bottom line shows number of target nodes found.} 
\label{tab:random}
\end{table}

%  \hline
%   random query probability    & positives found \\ 
%  \hline
%   (original) 0.0\%   & 709.5 $\pm$ 39.0 \\
%  2.5\%     & 714.4 $\pm$ 30.2 \\
%  5.0\%     & 709.6 $\pm$ 35.6 \\
%  7.5\%     & 714.1 $\pm$ 34.5 \\
%  10\%      & 722.8 $\pm$ 25.5 \\
%  15\%      & 719.0 $\pm$ 27.3 \\ 
%  20\%      & 721.2 $\pm$ 27.9 \\ 

%EWRLS
%741.5 $\pm$ 42.2
%731.5 $\pm$ 35.9
%743.7 $\pm$ 27.4
%731.1 $\pm$ 55.5
%729.8 $\pm$ 41.6
%731.1 $\pm$ 26.8
%724.7 $\pm$ 27.3
We observe that the performance does not improve significantly for $p \geq 2.5$\%, either because the diversity is not increasing in a way that
translates into performance improvements or because all gains are offset by the samples wasted when querying
nodes at random.

% the hopeless "undirected" diversity
Instead of querying uniformly at random, we could query nodes according to a probability distribution that concentrates
most of the mass on the top $k$ nodes w.r.t.\ model scores. 
We experimented with several ways of mapping scores to a probability distribution $P$. In particular,
we considered two classes of distributions:
\begin{itemize}
\item truncated geometric distribution ($0 < q < 1$):
\[
P(v) \propto (1-q)^{\pi(v)-1} q, \quad \textrm{and}
\]
\item truncated Zeta distribution ($r \geq 1$):
\[
P(v) \propto \pi(v)^{-r},
\]
\end{itemize}
where $\pi(v)$ is the rank of $v$ based on the scores given by the model to $v \in \mathcal{B}(t)$.
In each experiment, we set $q$ or $r$ at each step in one of nine ways:
\begin{enumerate}
\item Top 10 have $x\%$ of the probability mass; for $x \in \{70,90,99\}$.
\item Top 10\% nodes have $x\%$ of the probability mass; for $x \in \{90,99,99.9\}$.
\item Top $k(t) = \min\{10\times(1-t/T),1\}$ have $x\%$ of the probability mass; for $x \in \{70,90,99\}$.
\end{enumerate}
None of the mappings was able to substantially increase the search's performance.
In contrast to almost $20\%$ performance improvement seen by SVR under round-robin on CiteSeer at $T=1500$ (Fig.~\ref{fig:hitratio}),
mapping scores to a probability distribution increased the number of targets nodes found by at most~3\%.

\section{Evaluation of MAB algorithms applied to \ProbName}\label{app:mabs}

We experiment with representative algorithms of each of the following bandit classes:
\begin{itemize}
\item Stochastic Bandits: UCB1, Thompson Sampling (TS), $\epsilon$-greedy,
\item Adversarial Bandits: Exp3 \cite{Auer:2002hg},
\item Non-stationary stochastic bandits: Dynamic Thompson Sampling (DTS) \cite{Gupta:2011df},
\item Contextual Bandits: Exp4 \cite{Auer:2002hg} and Exp4.P \cite{Beygelzimer:2011wp}.
\end{itemize}

\begin{figure*}[h]
  \center
        \subfloat{\includegraphics[width=0.33\textwidth]{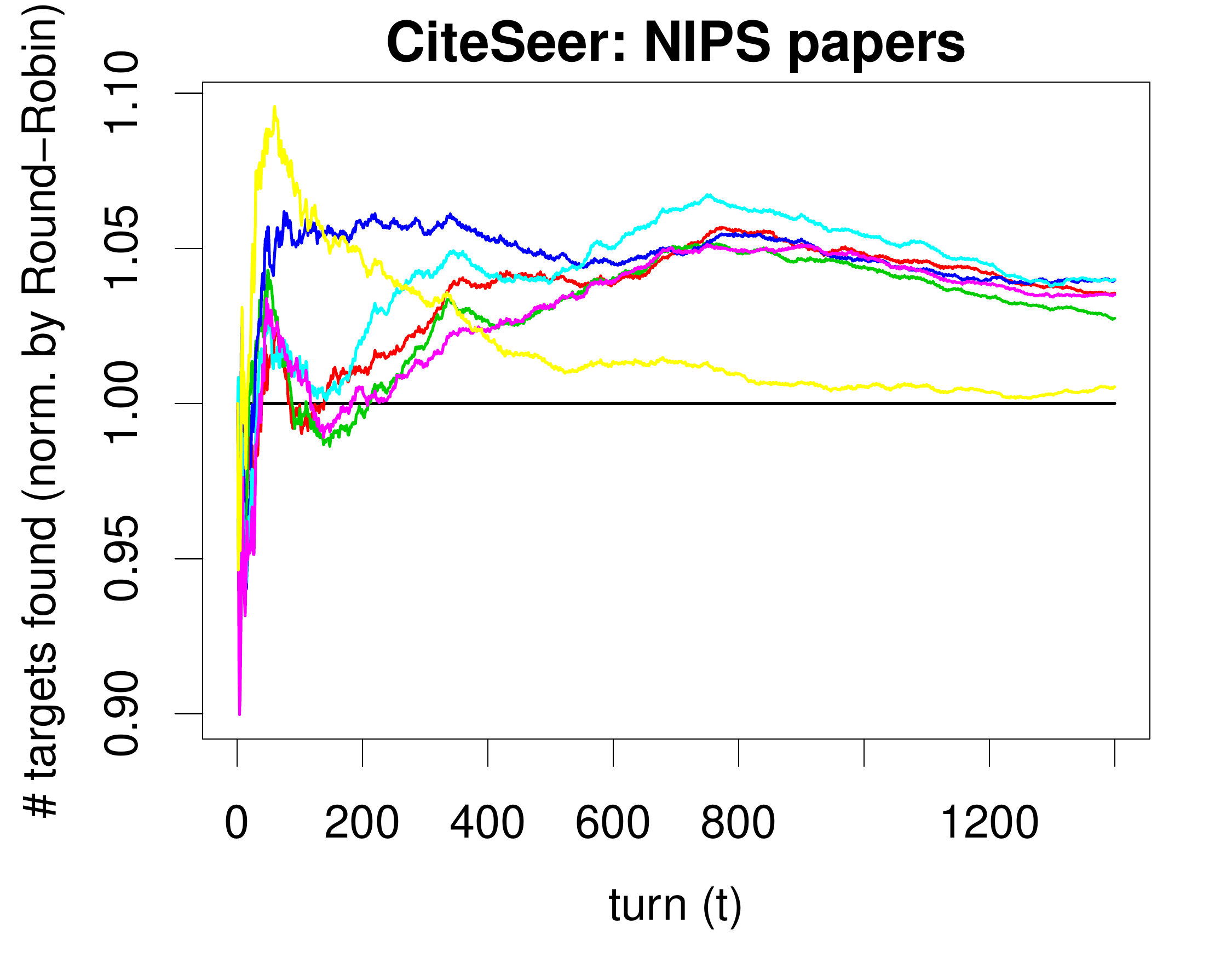}
     \label{fig:wikipedia_rr}}
    \subfloat{\includegraphics[width=0.33\textwidth]{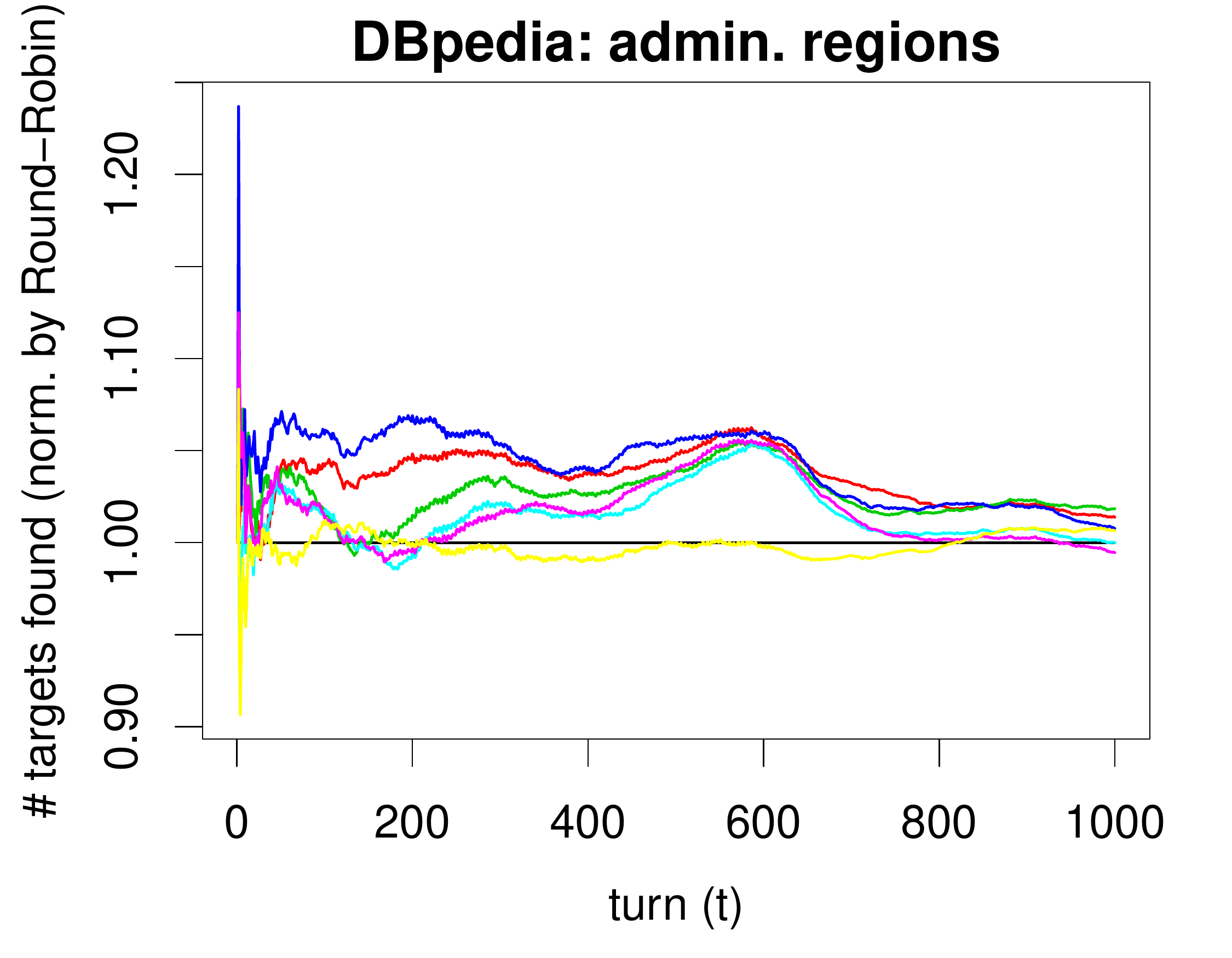}
    \label{fig:donors_rr}}
    \subfloat{\includegraphics[width=0.33\textwidth]{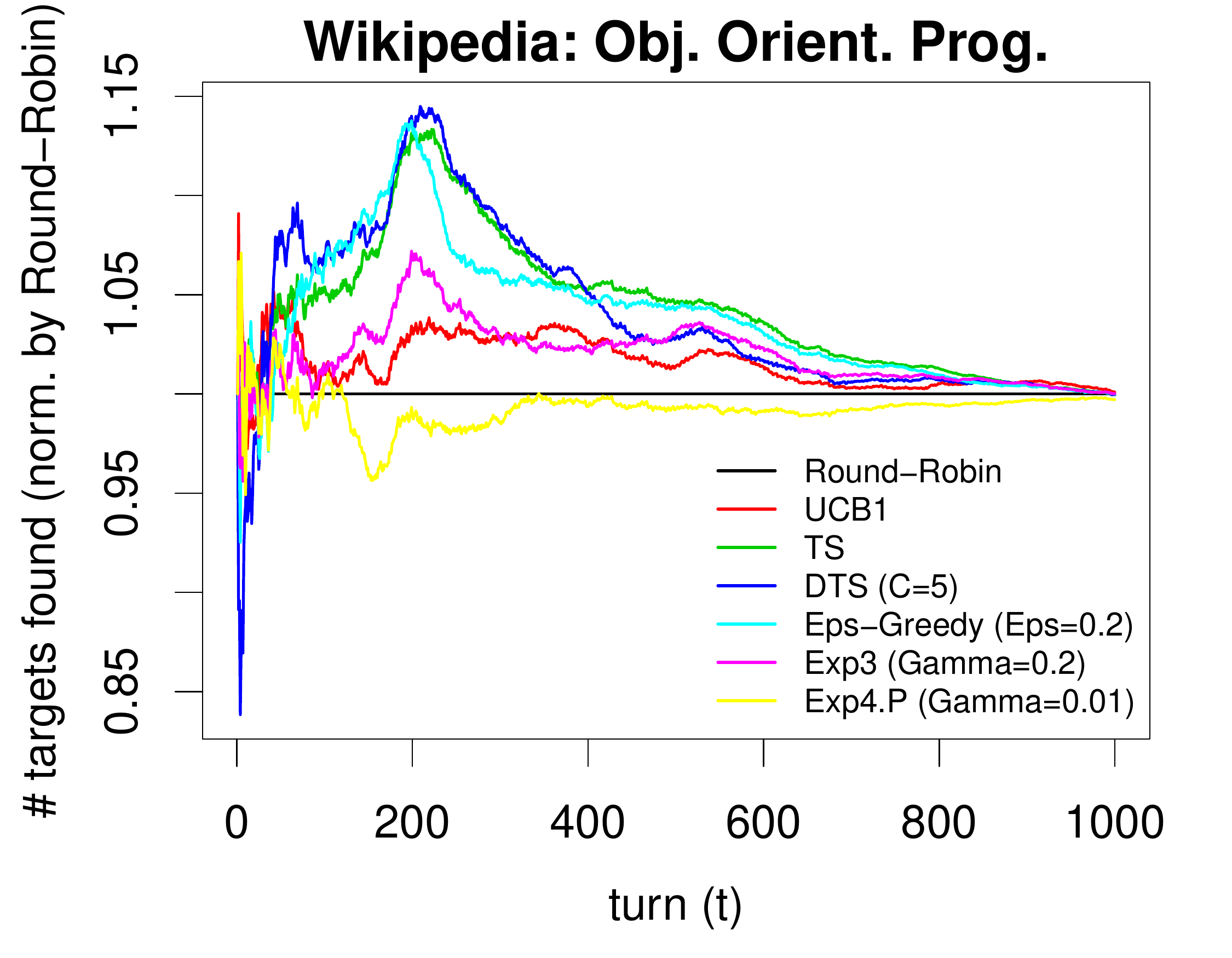}
     \label{fig:kickstarter_rr}}
\caption{Comparison between the best parameterizations of each MAB algorithm.}
\label{fig:mabs}
\end{figure*}

UCB1 and TS are parameter-free. For $\epsilon$-greedy, Exp3 and Exp4.P we set the probability of
uniformly random pulls, to $\epsilon \in \{0.10,0.20,0.50\}$, $\gamma \in \{0.10,0.20,0.50\}$ and $Kp_{\min} \in \{0.01,0.05,0.10,0.20,0.50\}$ (respectively).
We set parameter $\gamma$ in Exp4 as $K_p{\min}$ in Exp4.P.  
For DTS, we set the cap on the parameter sum $C \in \{5,10, 20,50\}$. Interestingly, for each MAB algorithm,
there was always one parameter value that outperformed all the others in almost all seven datasets. In Figure~\ref{fig:mabs}
we show three representative plots of the performance comparison between the best parameterizations of each MAB algorithm.
Since Exp4 was slightly outperformed by Exp4.P, Exp4 is not shown. These
results corroborate our expectations (Section~\ref{sec:d3ts}) that DTS would outperform other bandits in \probname problems.

\end{document}